\documentclass[fleqn,10pt]{wlscirep}
\usepackage[utf8]{inputenc}
\usepackage[T1]{fontenc}

\title{Iterative embedding and reweighting of complex networks reveals community structure}

\author[1]{Bianka Kov{\'a}cs}

\author[2,3]{Sadamori Kojaku}

\author[4,1,*]{Gergely Palla}

\author[2]{Santo Fortunato}

\affil[1]{Department of Biological Physics, Eötvös Lor{\'a}nd University, P{\'a}zm{\'a}ny P.\ stny.\ 1/A, Budapest, H-1117, Hungary}

\affil[2]{Luddy School of Informatics, Computing, and Engineering, Indiana University, 1015 East 11th Street, Bloomington, 47408, Indiana, USA}

\affil[3]{Department of Systems Science and Industrial Engineering, SUNY Binghamton, P.O. Box 6000, Binghamton, 13902, New York, USA}

\affil[4]{Health Services Management Training Centre, Semmelweis University, K{\'u}tv{\"o}lgyi {\'u}t 2., Budapest, H-1125, Hungary}

\affil[*]{gergely.palla@emk.semmelweis.hu}

\begin{abstract}
Graph embeddings learn the structure of networks and represent it in low-dimensional vector spaces. Community structure is one of the features that are recognized and reproduced by embeddings. We show that an iterative procedure, in which a graph is repeatedly embedded and its links are reweighted based on the geometric proximity between the nodes, reinforces intra-community links and weakens inter-community links, making the clusters of the initial network more visible and more easily detectable. The geometric separation between the communities can become so strong that even a very simple parsing of the links may recover the communities as isolated components with surprisingly high precision. 
Furthermore, when used as a pre-processing step, our embedding and reweighting procedure can improve the performance of traditional community detection algorithms. 
\end{abstract}

\DeclareMathAlphabet{\mathbfit}{OML}{cmm}{b}{it} 
\DeclareMathAlphabet{\pazocal}{OMS}{zplm}{m}{n}
\SetMathAlphabet{\pazocal}{bold}{OMS}{zplm}{b}{n}

\begin{document}
\flushbottom
\maketitle
\thispagestyle{empty}

\section*{Introduction}\label{sect:introduction}
Recent advances in machine learning have opened new productive research directions in the study of networks (or graphs). Graph embeddings are paradigmatic examples. They represent the structure of a graph via the geometric relations of a set of points arranged in a low-dimensional vector space, where the points are the network nodes and some features of the original network are preserved. Once the graph has been embedded, one can operate on the resulting spatial distribution of points by using the wealth of tools that are available in continuous metric spaces, in particular the possibility of computing distances between the points.

Graph embeddings have been instrumental in various graph data applications, including link prediction~\cite{Zhou_2011_link_prediction_survey,chenPMEProjectedMetric2018,kunegisLearningSpectralGraph2009,masrourBurstingFilterBubble2020}, node classification~\cite{Bhagat2011_node_classification,DeepWalk_2014,Deep_Network_Embedding_2016,node2vec,HOPE}, and community detection~\cite{Donath1973_spectral_clustering,Fiedler1973_spectral_clustering,Spielman_1996_spectral_clustering,Luxburg_2006_spectral_clustering,fortunato10,fortunato16,fortunato22,commSector_commDetMethod,clusteringOnHypEmb_criticalGap,coalescentEmbedding,barotCommunityDetectionUsing2021,LouvainAndLeidenWithEmbedding,commDetInEmbeddingsAndOtherThings,comparing4embsInCommDet,commDetWithEmb_dirNetworks,analogyBetweenHypEmbAndComms,commDetInEmbs_Santo,commDetInNeuralEmbs_Sadamori&Santo,communityEmbedding1,communityEmbedding2,neuralNetworkForEmbAndClustering_attributedNodes, deepGNNforEmbAndClustering_attributedNodes}.
Community detection is a pivotal task in network analysis because communities play key roles in the dynamics and functionality of networks~\cite{salathe2010dynamics,Dong2018ResilienceON,masuda2017random}. 
Communities are groups of nodes with a significant density of internal links, whereas the density of links connecting the groups to each other is comparatively lower. Since graph embedding methods typically place closely connected nodes in a network at nearby points in the embedding space, prominent communities are often embedded as compact, well-separated clusters~\cite{Luxburg_2006_spectral_clustering,Zhang2021ConsistencyOR}. These clusters can then be identified using data clustering techniques such as $k$-means clustering~\cite{kMeansClustering} or DBSCAN~\cite{DBSCAN}. Alternatively, the node proximity in the embedding can be used to facilitate network community detection algorithms by generating a good initial partition~\cite{LouvainAndLeidenWithEmbedding} or defining link weights~\cite{coalescentEmbedding}. Whether it is used for data clustering or enhancing network community detection algorithms, the applicability of graph embedding for the identification of communities depends on the ability of the embedding to project communities into distinct, compact clusters. This can be challenging, particularly when different communities are connected by many links. However, even if communities are not well separated in the network, embeddings can still capture node proximities, tending to place nodes within the same community closer together. This proximity information can be leveraged to refine the embedding, resulting in better-defined, compact community clusters that can be more easily identified using data clustering techniques.

We propose an iterative procedure, called \textit{Iterative Embedding and ReWeighting} (IERW), consisting of embedding the network and reweighting its links until a stable weighted graph configuration is reached. We find that, by utilizing information about node proximities derived from the embedding, we can obtain weighted networks in which the communities of the original graph are more and more pronounced over the iterations and easier to find. This effect can be so strong that it allows the recovery of communities by simply removing the longest links of the final weighted graph and identifying the connected components of the resulting network. This simple method is competitive with traditional community detection methods on synthetic graphs generated by the planted partition (PP) model~\cite{plantedPartitionModel} and can outperform them on the more realistic Lancichinetti--Fortunato--Radicchi (LFR) benchmark~\cite{LFRbenchmark}. Delivering link weights that strengthen the communities of the original network, IERW can also improve the performance of traditional community detection methods like Louvain~\cite{Louvain}, Infomap~\cite{Infomap} or label propagation~\cite{alabprop}. 

Formerly, an iterative embedding method has been proposed in the field of graph neural networks, where both the graph structure and the embedding are learned in an iterative manner, aiming for a better representation~\cite{Chen_iterative_graph_neural_net}. In parallel, an iteration of node2vec embedding~\cite{node2vec} using $k$-means clustering~\cite{kMeansClustering} cost regularization has been also proposed~\cite{Makarov_iterative_embedd_and_clust}, whereas in an alternative approach, specifically tailored for hyperbolic embedding based on the random hyperbolic graph~\cite{hyperGeomBasics,Mercator_H2model}, the model likelihood was regularized iteratively by taking into account also the communities~\cite{Ye_community_preserving_iterative_hyp_embedding}. Our work provides a more general framework, allowing the inclusion of any embedding method in general. In the present study, we apply both Euclidean and hyperbolic embedding algorithms, all leading to similar results at the qualitative level. 

\section*{Results}\label{sect:results}

\subsection*{Iterative embedding and reweighting}

\begin{figure}[!h]
    \centering
    \includegraphics[width=1.0\textwidth]{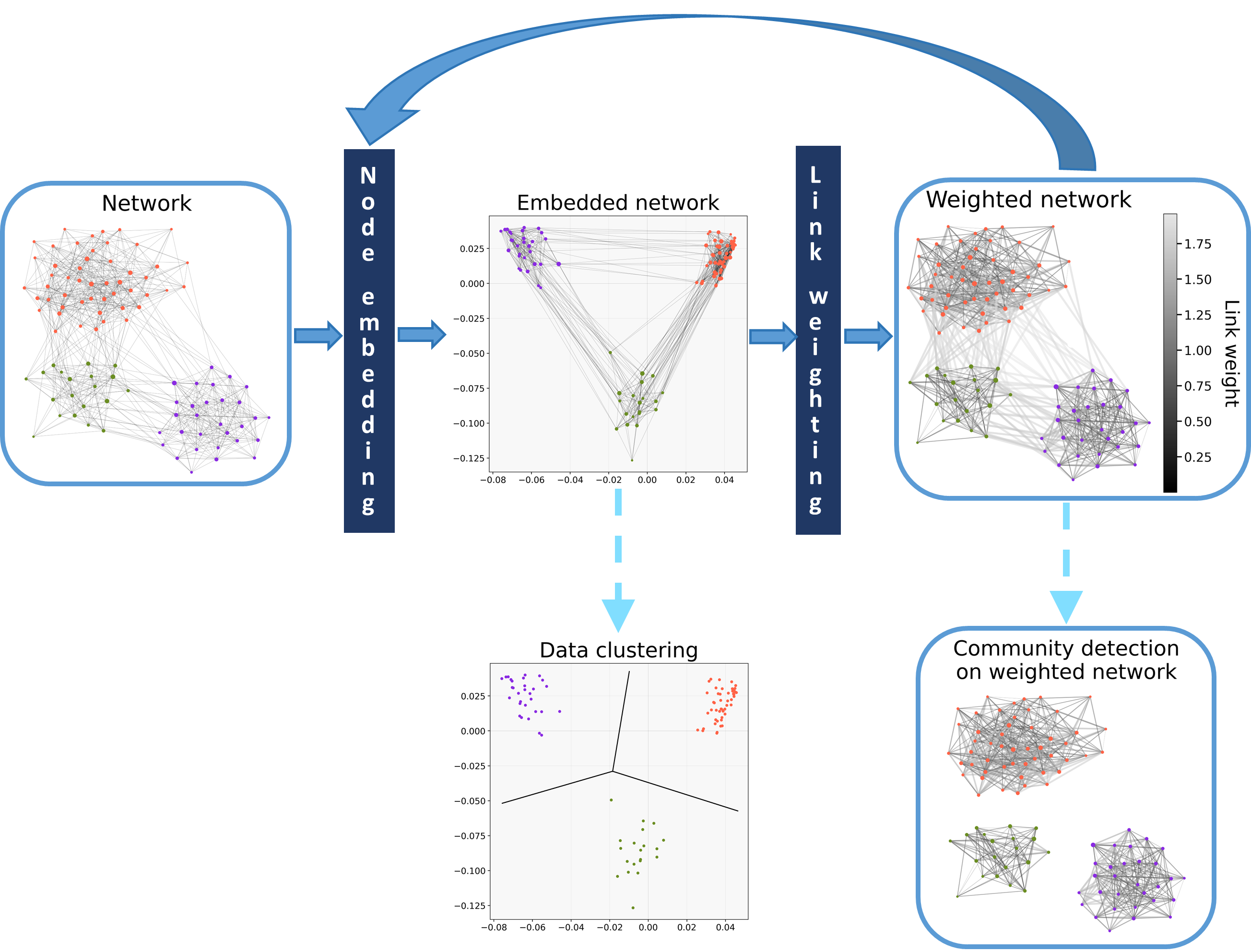}
    \caption{ {\bf Flowchart of the Iterative Embedding and ReWeighting process.} IERW embeds a network into a vector space, where nodes belonging to the same communities are closer to each other compared to nodes from different communities. Then, IERW generates a weighted network with the same sets of nodes and edges, where the edge weights reflect the angular relations of the network nodes in the embedding space. Repeating these two steps, IERW iteratively embeds a weighted network and reweights its links until the variation in the average edge weight within one iteration falls below a specified threshold. Finally, the communities can be identified using data clustering or network community detection algorithms. The example network was generated by the stochastic block model and embedded with Laplacian Eigenmaps on the Euclidean plane. The coloring of the nodes indicates the block memberships assigned by the stochastic block model.}
    \label{fig:flowchart}
\end{figure}

Given an embedding that can form dense spatial clusters from nodes that are strongly connected to each other, it can be expected that when the cohesiveness within the network communities and the separation between them are enhanced via some link weights, then a repeated embedding can further increase the density of the initial spatial clusters. 
Following this concept, as it is shown in Fig.~\ref{fig:flowchart}, the proposed Iterative Embedding and ReWeighting (IERW) process repeatedly arranges the network nodes in a vector space according to the topological relations between them and assigns weights to the links of the network in accordance with the geometric relations between the nodes in the previous embedding. This framework provides two opportunities for community detection: one can either use standard data clustering methods on the spatial node arrangements generated by the embedding steps, or utilize both the network topology and the geometric relations between the nodes by applying a community detection method on the weighted networks obtained from the link weighting steps.

While our IERW framework is agnostic to the method applied for network embedding, we illustrate the effectiveness of the iterative embedding by focusing on four embedding algorithms (described in the Methods section): Laplacian Eigenmaps (LE)~\cite{LE}, TRansformation of EXponential shortest Path lengths to hyperbolIC measures (TREXPIC)~\cite{TREXPIC}, Isomap (ISO)~\cite{Isomap} and node2vec~\cite{node2vec}. All the applied embedding methods are capable of embedding connected, possibly weighted undirected networks without self-loops and parallel edges in either Euclidean (LE, ISO, node2vec) or hyperbolic (TREXPIC) spaces of any number of dimensions $d$. While LE, TREXPIC and ISO are dimensional reduction techniques based on matrix factorization, 
in node2vec a neural network creates embeddings based on random walks performed along the network.

As it is detailed in the Methods section, two of the considered methods, namely LE and TREXPIC build on relatively fast-changing, exponential measures of the topological proximity and distance between the network nodes. Following this idea, in order to emphasize the differences between the connectedness of different node pairs in the case of ISO as well, we created a modified version of this embedding method by inserting an exponentialization step into the algorithm. Similarly, we also included exponentialization in the iteration of node2vec, where we left the embedding algorithm itself unaltered but chose an exponential link weight function in IERW. The positive effect of introducing exponentialization in ISO and node2vec is demonstrated in Sect.~\ref{sect:expVSnotExp} of the Supplementary Information. In all figures appearing in this paper, we utilised the exponentialization in both ISO and node2vec. Note that the exponentialization step has a tunable constant $t>0$ in the case of all four embedding methods. We did not search for its optimal value in each task individually but used the default setting in all of our measurements. Therefore, our results achieved with IERW may not be the best possible outcomes and there may be room for improvement. The effect of changing $t$ in the exponentialization step of the different embedding methods is examined in Sect.~\ref{sect:embParams} of the Supplementary Information.

A crucial step of IERW is the calculation of the link weights based on the positions of the connected nodes in the previous embedding. It is important to bear in mind that the different embedding methods may need different types of link weights as input. Traditionally, in network science link weights represent the intensity or strength of the connection, where a high weight value refers to a strong, close relation between the given node pair. However, some of the embedding methods originate from algorithms initially designed to provide low-dimensional approximations of distances in high-dimensional point clouds, where a high value associated to a node pair refers to a high distance and therefore, presumably a weak connection or a distant relation. Among the embedding methods used in this paper, LE, TREXPIC and ISO expect such distance-like link weights when encountering a weighted link list as an input. In contrast, node2vec expects proximity-like link weights, matching the traditional weight definition in network science.

All the four examined embedding methods tend to place the nodes within the same communities at rather similar angular coordinates, i.e. at small angular distances $\Delta\theta$ from each other. For this reason, to make the embeddings gradually more focused on the community structure, we always defined the link weights in IERW based on the angular relations between the connected nodes. Since cosine distance and cosine proximity are both well-known measures of the angular relations of network nodes, we built our link weighting formulas in IERW on $\cos(\Delta\theta)$. This can be easily calculated for the $d$-dimensional Cartesian position vectors $\underline{y}_i$ and $\underline{y}_j$ of nodes $i$ and $j$ in both the Euclidean and the hyperbolic embedding space as 
\begin{equation}
    \cos(\Delta\theta_{ij})=\frac{\underline{y}_i\cdot\underline{y}_j}{\|\underline{y}_i\|\,\|\underline{y}_j\|}
    \label{eq:cosOfAngDist}
\end{equation}
from the dot product $\underline{y}_i\cdot\underline{y}_j=\sum_{\ell=1}^d\underline{y}_i(\ell)\,\underline{y}_j(\ell)$ and the Euclidean norms {$\|\underline{y}_i\|=\sqrt{\sum_{\ell=1}^d \underline{y}_i(\ell)^2}$} and $\|\underline{y}_j\|=\sqrt{\sum_{\ell=1}^d \underline{y}_j(\ell)^2}$. The exact definition of the link weighting formula applied in IERW is given in the Methods section for each embedding method.

Besides the link weights, we also have to specify the number of dimensions $d$ of the embedding space and a stopping criterion for the iteration to make the IERW framework completed. In the case of the matrix factorization methods (LE, TREXPIC and ISO), we aimed for an embedding dimension $d$ equal to $d=C-1$, where $C$ denotes the supposed number of communities in the network, which we determined from the eigengap of a normalized graph Laplacian (see Methods). Node2vec, however, often works better with a large $d$ in practice due to the nature of the training algorithm. More specifically, node2vec is trained with the stochastic gradient descent algorithm, which regularizes node2vec and prevents it from overfitting~\cite{smith2021origin}. 
For this reason, we simply used node2vec with a fixed value of $d=64$, corresponding to one of the standard choices in the literature. According to our measurements presented in Sect.~\ref{subsect:measuringTheEffectOfEmbParams} of the Supplementary Information, while LE, TREXPIC and ISO indeed seem to require a rather specific number of embedding dimensions, the performance of node2vec shows comparatively weak dependence on the value of $d$.

Finally, the stopping criterion for IERW was based on monitoring the relative change in the average link weight $\bar{w}$ between subsequent iterations, and the process was terminated when this quantity dropped below a certain threshold, namely when we reached
\begin{equation}
    \frac{\lvert \bar{w}_{\mathrm{current}}-\bar{w}_{\mathrm{previous}} \rvert}{\bar{w}_{\mathrm{current}}}\leq0.001.
    \label{eq:stoppingCriterion}
\end{equation}
Note that we stopped the iteration process after the $20$th iteration even if the stopping criterion in Eq.~(\ref{eq:stoppingCriterion}) has not been fulfilled yet. 

\begin{figure}[!h]
    \centering
    \includegraphics[width=1.0\textwidth]{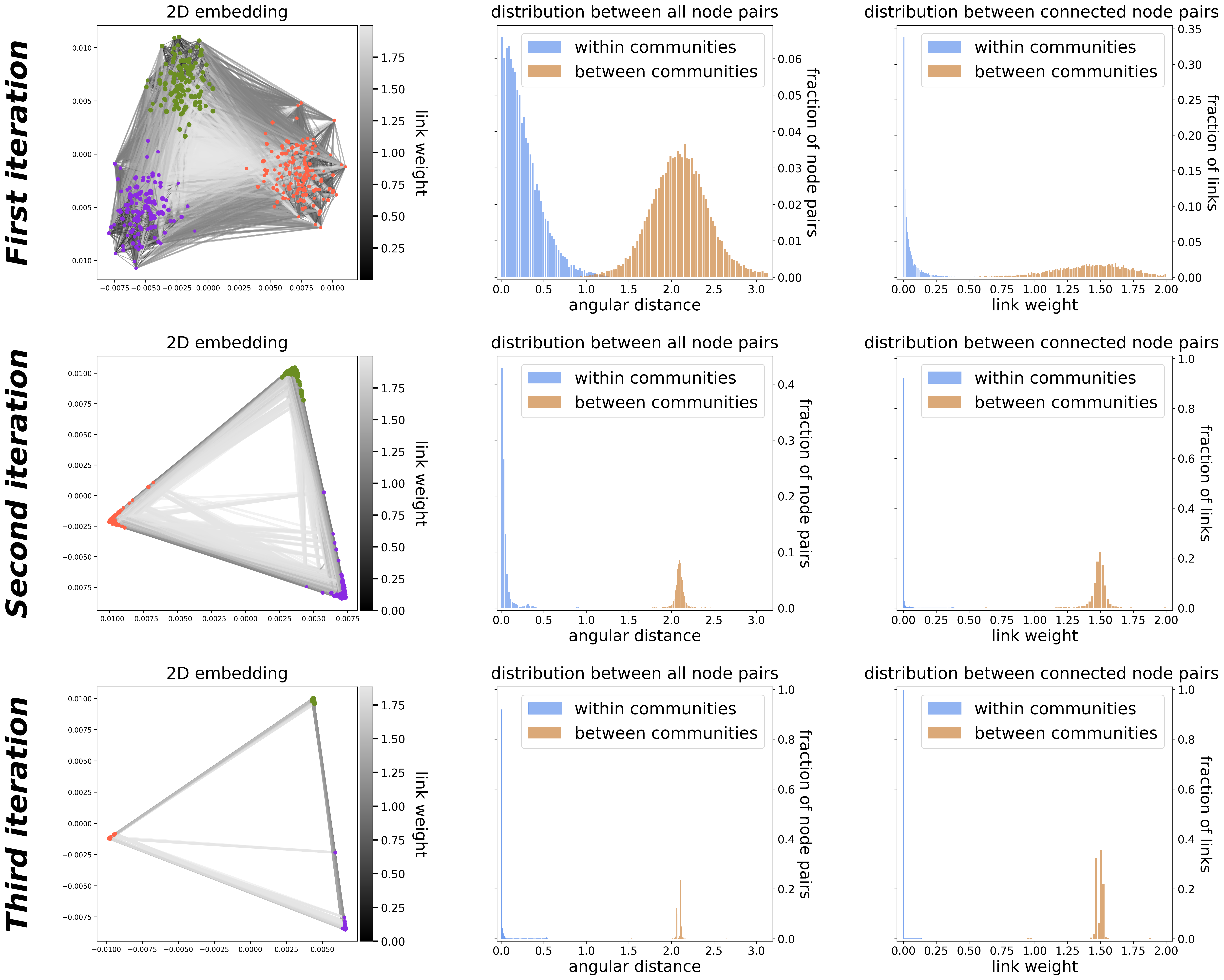}
    \caption{ {\bf Example of IERW}. A network with three communities built by the stochastic block model was embedded three times in the $2$-dimensional Euclidean space with Laplacian Eigenmaps. Each row of panels corresponds to one iteration. Initially, all the link weights were $1$, and we updated the weights after each embedding using the angular distances $\Delta\theta_{ij}$ as $w_{ij}=1-\cos(\Delta\theta_{ij})$. The left column of panels shows the embeddings, denoting the smaller link weights (that indicate smaller angular distances, and thus, stronger connections) at the end of the given iteration with darker and narrower lines, and coloring the network nodes according to the planted blocks. The column in the middle shows the distribution of the angular distances between all the node pairs in the embedding of the given iteration, while the right column shows the distribution of the link weights of the network.
    }
    \label{fig:iterationExample}
\end{figure}

To demonstrate how IERW works, Fig.~\ref{fig:iterationExample} shows three iterations using LE, performed on a network generated by the stochastic block model (SBM)~\cite{simplestSBMarticle} with three communities of size $\lvert \pazocal{A}\rvert =150$ (orange), $\lvert \pazocal{B}\rvert=130$ (purple) and ${\lvert \pazocal{C}\rvert=120}$ (green). In the SBM, the link probability between two nodes only depends on their respective memberships. For three communities these probabilities thus fill a $3\times 3$ stochastic block matrix $\mathbfit{M}$, which in our case is
\begin{equation}
\begin{array}{@{}c@{\hspace{1ex}}c@{}}
 & \textit{$\pazocal{A}$\,\,\,\,\,\,\,\,\,$\pazocal{B}$\,\,\,\,\,\,\,\,\,$\pazocal{C}$} \\[0.5ex]
\mathbfit{M} \,\,=\,\,\, \rotatebox[origin=c]{90}{\textit{$\pazocal{C}$\,\,\,$\pazocal{B}$\,\,\,$\pazocal{A}$}} &
 \begin{bmatrix}
  0.30 & 0.10 & 0.15\\
  0.10 & 0.35 & 0.20\\
  0.15 & 0.20 & 0.40
 \end{bmatrix}.
\end{array}
\label{eq:SBMconnProbMatrix}
\end{equation}
According to Fig.~\ref{fig:iterationExample}, IERW turns the communities into more and more concentrated spatial clusters. Consequently, the distribution of the angular distances between all the node pairs (middle column) and also between the connected node pairs (right column) split into two peaks each with increasing separation. One peak corresponds to the node pairs of the same community (blue) whereas the other refers to the node pairs in different communities (orange).
 
\subsection*{Angular separation of communities in iteratively embedded networks}
\label{subsect:results_angSepRates}

We applied IERW to synthetic networks generated by the planted partition (PP) model~\cite{plantedPartitionModel} or the Lancichinetti--Fortunato--Radicchi (LFR) model~\cite{LFRbenchmark}. A key advantage of these generative models is that they enable the definition of communities with tunable internal and external link densities, allowing to control the difficulty of the network clustering problem through the adjustment of the mixing parameter $\mu$, which corresponds to the average fraction of neighbors of one node belonging to communities different from the one of the node. Details on the synthetic network generation are provided in the Methods section.

\begin{figure}[!h]
    \centering
    \includegraphics[width=1.0\textwidth]{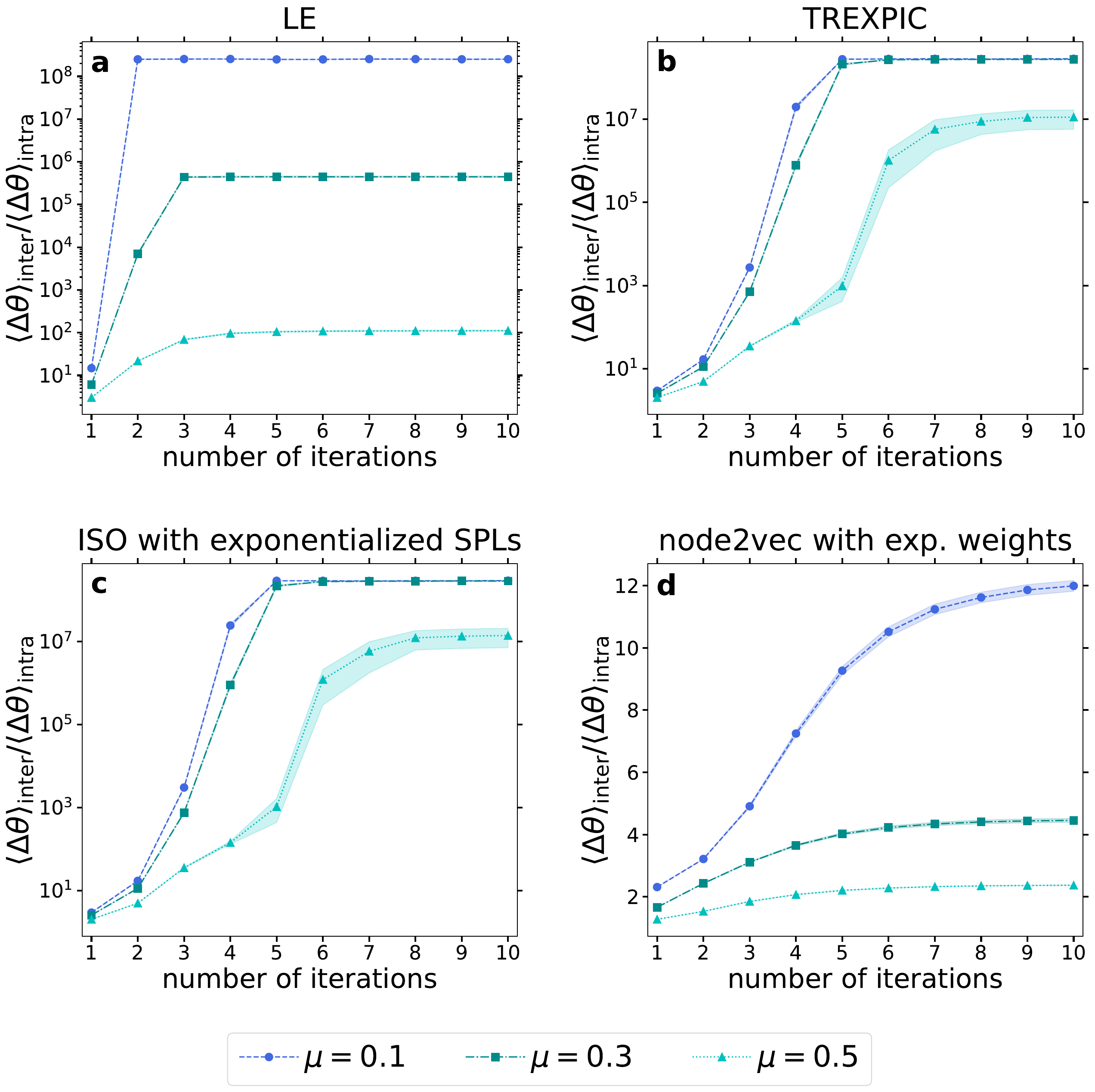}
    \caption{ {\bf IERW increases angular separation of communities in networks generated by the PP model.} 
    We plot the ratio between the average angular distance of all possible node pairs in different communities and in the same community as a function of the number of IERW iterations for LE (\textbf{a}), TREXPIC (\textbf{b}), ISO with exponentialized shortest path lengths (\textbf{c}) and node2vec (\textbf{d}). Curves of different colors correspond to different values of the mixing parameter $\mu$. Each depicted data point was obtained by averaging the results over $100$ different network realizations, and the shaded areas show the standard error of the mean.}
    \label{fig:angSepRate_PP}
\end{figure}

In Fig.~\ref{fig:angSepRate_PP}, we show the ratio between the average inter-community angular distance $\langle \Delta\theta \rangle_{\rm inter}$ (i.e., the average of the angular distances over all the node pairs of different communities) and the average intra-community angular distance $\langle \Delta\theta \rangle_{\rm intra}$ (i.e., the average of the angular distances over all the node pairs belonging to the same community) as a function of the number of IERW iterations performed for networks generated by the PP model. According to the figure, the $\langle \Delta\theta \rangle_{\rm inter}/\langle \Delta\theta \rangle_{\rm intra}$ ratio starts to increase over the iterations and then saturates for all the four studied embedding methods, reaching in some cases extremely high values, which indicates a strong separation between the planted communities in the embedding space. Naturally, when the mixing parameter $\mu$ is only $0.1$, the angular separation ratio $\langle \Delta\theta \rangle_{\rm inter}/\langle \Delta\theta \rangle_{\rm intra}$ is higher compared to the case of moderate mixing between the communities at $\mu=0.3$, that in turn surpasses in every iteration the results observed for the relatively strong mixing of $\mu=0.5$, where nodes have roughly the same number of internal and external neighbors. Nevertheless, the curves of the angular separation ratio are increasing as a function of the number of iterations even at $\mu=0.5$, indicating that our iterative embedding framework helps in separating the planted communities in the embedding space. 

\begin{figure}[!h]
    \centering
    \includegraphics[width=1.0\textwidth]{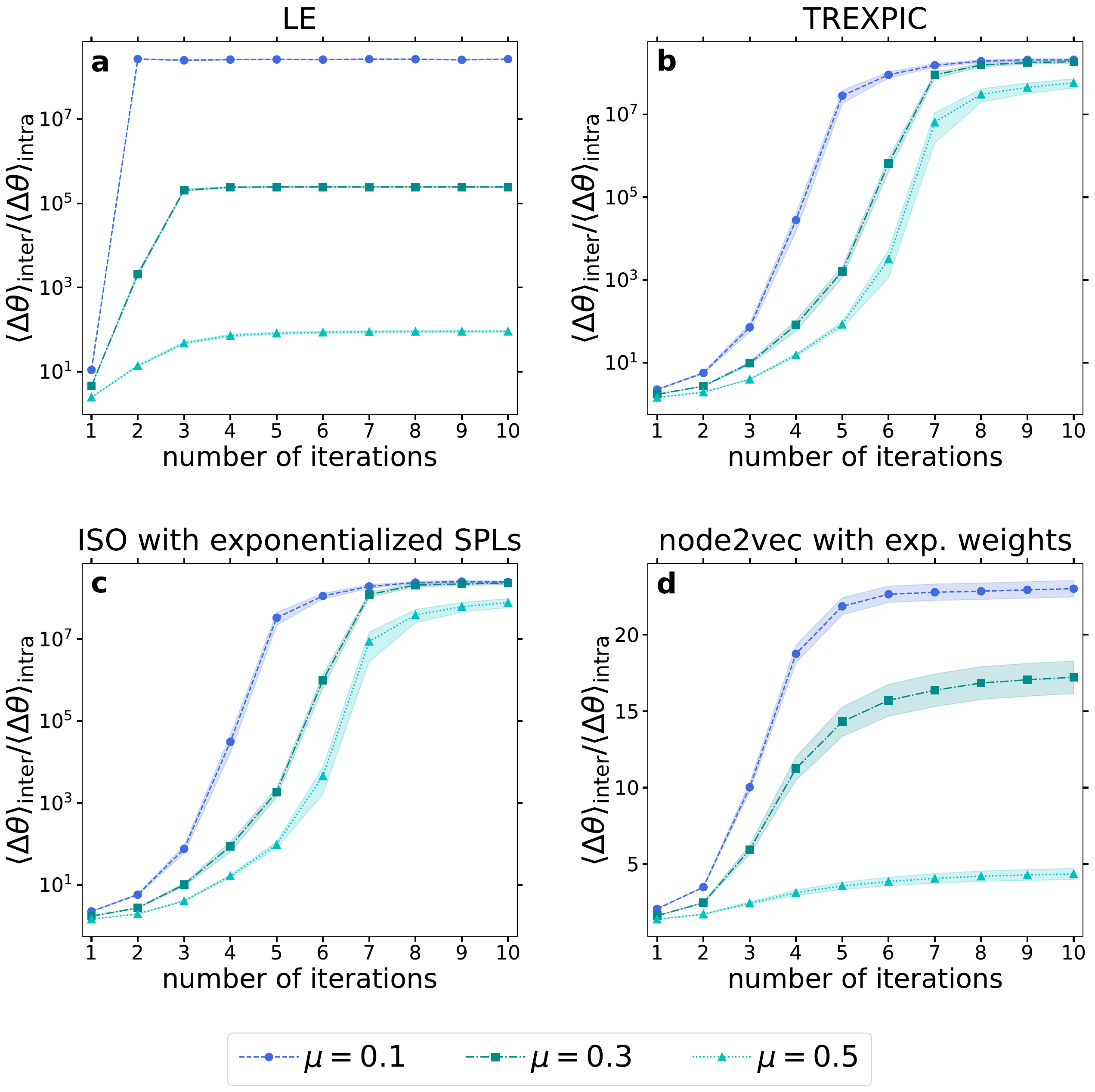}
    \caption{ {\bf IERW increases angular separation of communities in networks generated by the LFR benchmark.} 
    We plot the ratio between the average angular distance of all possible node pairs in different communities and in the same community as a function of the number of IERW iterations for LE (\textbf{a}), TREXPIC (\textbf{b}), ISO with exponentialized shortest path lengths (\textbf{c}) and node2vec (\textbf{d}). Curves of different colors correspond to different values of the mixing parameter $\mu$. Each depicted data point was obtained by averaging the results over $100$ different network realizations, and the shaded areas show the standard error of the mean.}
    \label{fig:angSepRate_LFR}
\end{figure}

In Fig.~\ref{fig:angSepRate_LFR}, we display the results for the angular separation of planted communities in LFR networks. The qualitative behaviour of the $\langle \Delta\theta \rangle_{\rm inter}/\langle \Delta\theta \rangle_{\rm intra}$ ratio is quite similar to that in Fig.~\ref{fig:angSepRate_PP}: 
the angular separation ratio starts with an increasing trend and then saturates as a function of the number of IERW iterations. The lower the $\mu$ value, the higher the saturated ratio. As in Fig.~\ref{fig:angSepRate_PP}, the actual value of the angular separation ratio can grow even above $\langle \Delta\theta \rangle_{\rm inter}/\langle \Delta\theta \rangle_{\rm intra}=10^7$.


\subsection*{Separation of communities via weight thresholding}
\label{subsect:results_commDetWithThreshold}

To give further perspective on the communities' strong separation resulting from our framework, here we show that even a clearly sub-optimal, overly simplistic community extraction method can provide surprisingly good results when exploiting the geometric information encoded in the link weights at the end of the IERW process. The basic idea is to set a threshold aiming to separate the links that connect members of the same community from those between nodes of different communities. 
By deleting the links on one side of the threshold---those that are suspected to connect different communities---, the network falls apart into disconnected components that we may identify as the detected communities. Since, as it is illustrated by Fig.~\ref{fig:iterationExample}, a rather large gap can emerge between the weight of intra- and inter-community links during IERW, this simple weight thresholding strategy (detailed in the Methods section) can actually work effectively under optimal circumstances. 
Note that the iteration of the embedding is indeed necessary to make the weight thresholding work as the weight thresholding after a single embedding yields poor community detection performance (see Sect.~\ref{sect:weightThreshold_firstVSfinalIter} of the Supplementary Information).

\begin{figure}[!h]
    \centering
    \includegraphics[width=1.0\textwidth]{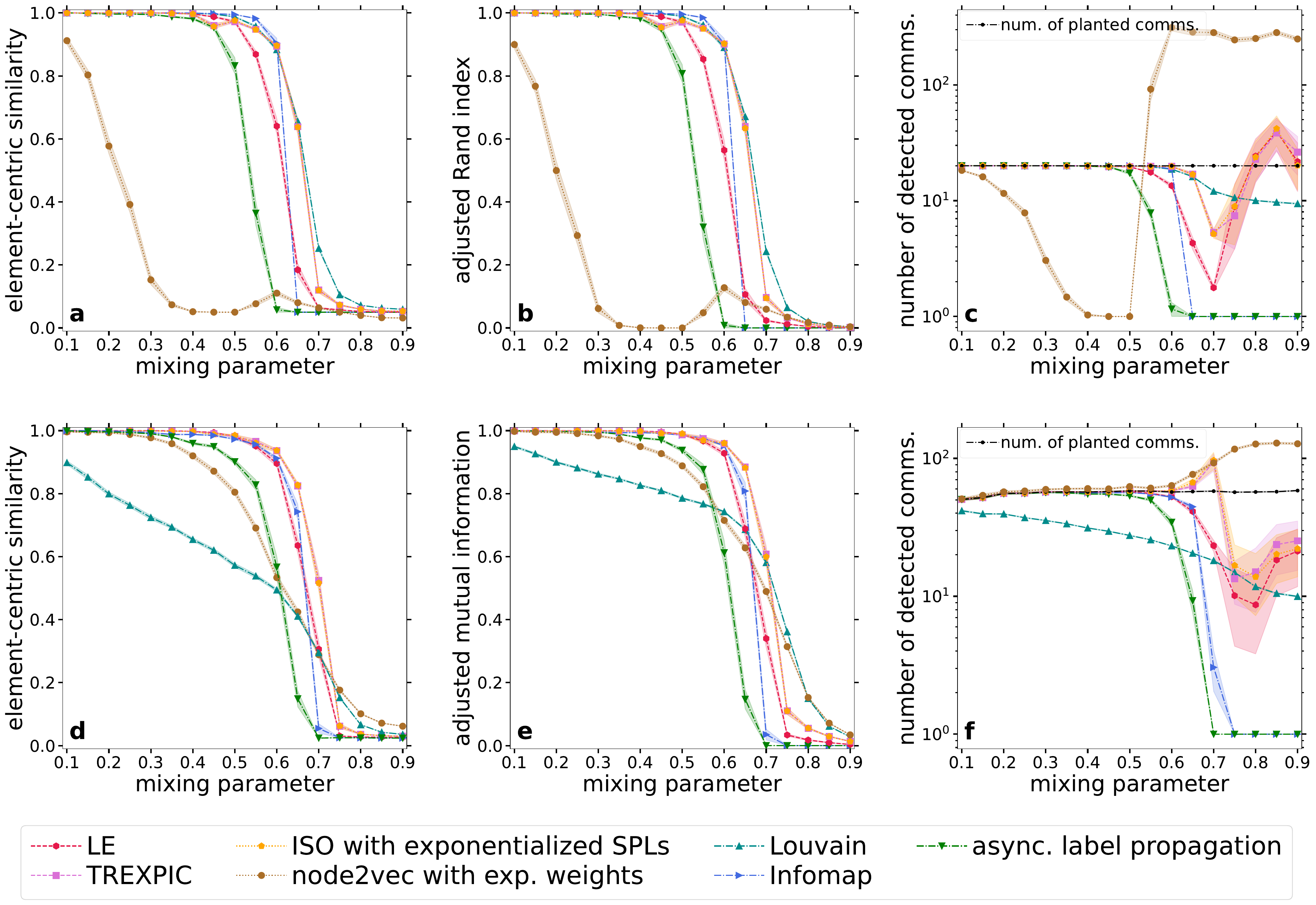}
    \caption{ 
    {\bf Extracting communities via weight thresholding the network yielded by IERW.} Panels \textbf{a}, \textbf{b} and \textbf{c} refer to input networks generated by the PP model, while panels \textbf{d}, \textbf{e} and \textbf{f} deal with input networks obtained from the LFR benchmark. As a reference, the three dash-dotted lines show the results achieved by traditional network community detection methods on the initial unweighted graphs: Louvain (dark cyan upward-pointing triangles), Infomap (blue right-pointing triangles) and asynchronous label propagation (green downward-pointing triangles). The other four colored lines illustrate the results for a simple weight thresholding that we applied on the final weighted networks obtained from IERW with LE (red hexagons), TREXPIC (purple squares), ISO with exponentialized shortest path lengths (orange pentagons) and node2vec with exponentialized link weights (brown circles). We performed the community detection with all the methods only once for each network. Each displayed data point corresponds to a result averaged over $100$ networks, and the error bars indicate the standard error of the mean.}
    \label{fig:commDetWithWeightThreshold}
\end{figure}

Although the applied weight thresholding approach is rather crude, thanks to the large angular separation that IERW achieves between the communities, it can still yield results comparable in quality to state-of-the-art community-finding methods. In Fig.~\ref{fig:commDetWithWeightThreshold}, we compare the performance of the weight thresholding with that of three commonly used, well-established network community detection methods. Even though all three methods are able to take into account link weights, in the case of Fig.~\ref{fig:commDetWithWeightThreshold} we applied them on the original, unweighted test graphs and not on the weighted versions obtained from the IERW process. 

First, we used the Louvain algorithm~\cite{Louvain,Louvain_code}, performing a heuristic maximization of the well-known modularity by Newman and Girvan~\cite{Newman_modularity_original,weightedModularityDef}, which compares the observed internal link density of the communities to its expected value. Though Louvain can unfold a hierarchical community structure (with nested modules and submodules), we always considered the top-level community structure, i.e. the one with the highest modularity. 

Besides, we applied the Infomap algorithm~\cite{Infomap,Infomap_code}, which relies on a heuristic minimization of the so-called map equation~\cite{Infomap}. It assumes that communities are regions of a network within which random walkers spend a relatively long time, and searches for the community structure that is the best for compressing the description (i.e., the code length) of random walk trajectories along the network. Infomap, just like Louvain, can create a hierarchy of network partitions; here we considered the lowest hierarchical level, yielding the shortest description length. 

Finally, we used the asynchronous label propagation algorithm~\cite{alabprop,alabprop_code}, which does not aim at the optimization of any predefined measure but simulates the diffusion of the nodes' community labels along the links, regularly updating the community membership of each node following the current majority of the neighboring nodes, expecting that eventually a consensus on a unique label becomes established within densely connected groups of network nodes. Following the suggestion in Ref.~\cite{alabprop}, we completed the label propagation process by separating groups of nodes that ended up with the same label but were not connected to each other.

The PP graphs (top row of panels) and the LFR networks (bottom row of panels) studied in Fig.~\ref{fig:commDetWithWeightThreshold} are the same as in Figs.~\ref{fig:angSepRate_PP} and \ref{fig:angSepRate_LFR}, respectively. The network generation process is detailed in the Methods section. To evaluate the performance of the examined community detection methods, we measured the number of detected communities (right column of Fig.~\ref{fig:commDetWithWeightThreshold}), as well as different similarity scores (left and middle columns of Fig.~\ref{fig:commDetWithWeightThreshold}) between the planted and the detected community structures.

First, we calculated the element-centric similarity (ECS)~\cite{elCentSim,elCentSimCode} between the detected and planted partitions (Fig.~\ref{fig:commDetWithWeightThreshold}a,d), which is a measure 
comparing node-node transition probabilities in random walks performed along the two graphs of cluster-induced (i.e., groupmate) relationships derived from the two partitions. ECS has its maximum of $1$ for identical partitions and decreases as the similarity between the compared divisions declines. Note that the expected value of ECS when inputting two random partitions having an equal number of groups and equal group sizes is not set to $0$~\cite{commDetInNeuralEmbs_Sadamori&Santo}. Furthermore, the 
only tunable parameter of the method for non-hierarchical clusterings is given by the restart probability of the random walks, but it does not have any effect in the case of hard partitions~\cite{commDetInNeuralEmbs_Sadamori&Santo}, so in our measurements we simply used its default value. 

Besides the ECS, following the suggestions of Ref.~\cite{ARIvsAMI}, we used the adjusted Rand index (ARI)~\cite{ARI1,ARI2,ARI3,ARI_code} for the PP networks (Fig.~\ref{fig:commDetWithWeightThreshold}b), where the group sizes in the ground truth clustering were equal, and the adjusted mutual information (AMI)~\cite{AMI1,AMI2,AMI3,AMI_code} for the LFR networks (Fig.~\ref{fig:commDetWithWeightThreshold}e), where the ground truth partition was unbalanced with respect to the group sizes, i.e. strongly different community sizes occurred. Both ARI and AMI take the value of $1$ in the case of perfect agreement between two partitions, and (being corrected or adjusted for the agreement emerging only by chance) the value of $0$ on expectation when comparing 
random partitions having the same number of communities and the same community sizes. 
ARI and AMI can decrease even below $0$ if the considered two clusterings differ 
to a large extent. While ARI is a pair-counting similarity measure that relies on the number of node pairs being groupmates or belonging to different groups in both the planted and the detected community structures, 
AMI is an information-theoretic quantity operating with the community membership probabilities of a randomly chosen node, which are calculated based on the relative size of the communities and the overlaps between the groups from the different partitions. 
Though there are several different possibilities for the normalization in the AMI formula, we always normalized with the maximum of the Shannon entropies associated with the two partitions to be compared.

In the case of the PP model, the community-finding performance of the weight thresholding based on iterated node2vec is poor according to both ECS (Fig.~\ref{fig:commDetWithWeightThreshold}a) and ARI (Fig.~\ref{fig:commDetWithWeightThreshold}b). 
In the meantime, the similarity scores achieved using IERW in the case of TREXPIC or ISO with exponentialized shortest path lengths are very close to that of Infomap and Louvain in Fig.~\ref{fig:commDetWithWeightThreshold}a,b. The results based on iterated LE fall slightly behind, although they still surpass the scores of asynchronous label propagation. 

In the case of the LFR benchmark, the results for the weight thresholding based on IERW using both TREXPIC and ISO with exponentialized shortest path lengths slightly exceed that of even Infomap (Fig.~\ref{fig:commDetWithWeightThreshold}d,e), which is followed closely by the results achieved using iterated LE. Asynchronous label propagation falls somewhat behind similarly to the PP case, but here it is followed relatively closely by the results based on iterated node2vec, which in turn surpasses Louvain. Louvain has a poor performance on LFR graphs due to the resolution limit of modularity maximization~\cite{resolution_limit_PNAS}.

\subsection*{Facilitating traditional community detection methods with iterative embedding}
\label{subsect:results_commDetWithUsualMethods}

\begin{figure}[!h]
    \centering
    \includegraphics[width=1.0\textwidth]{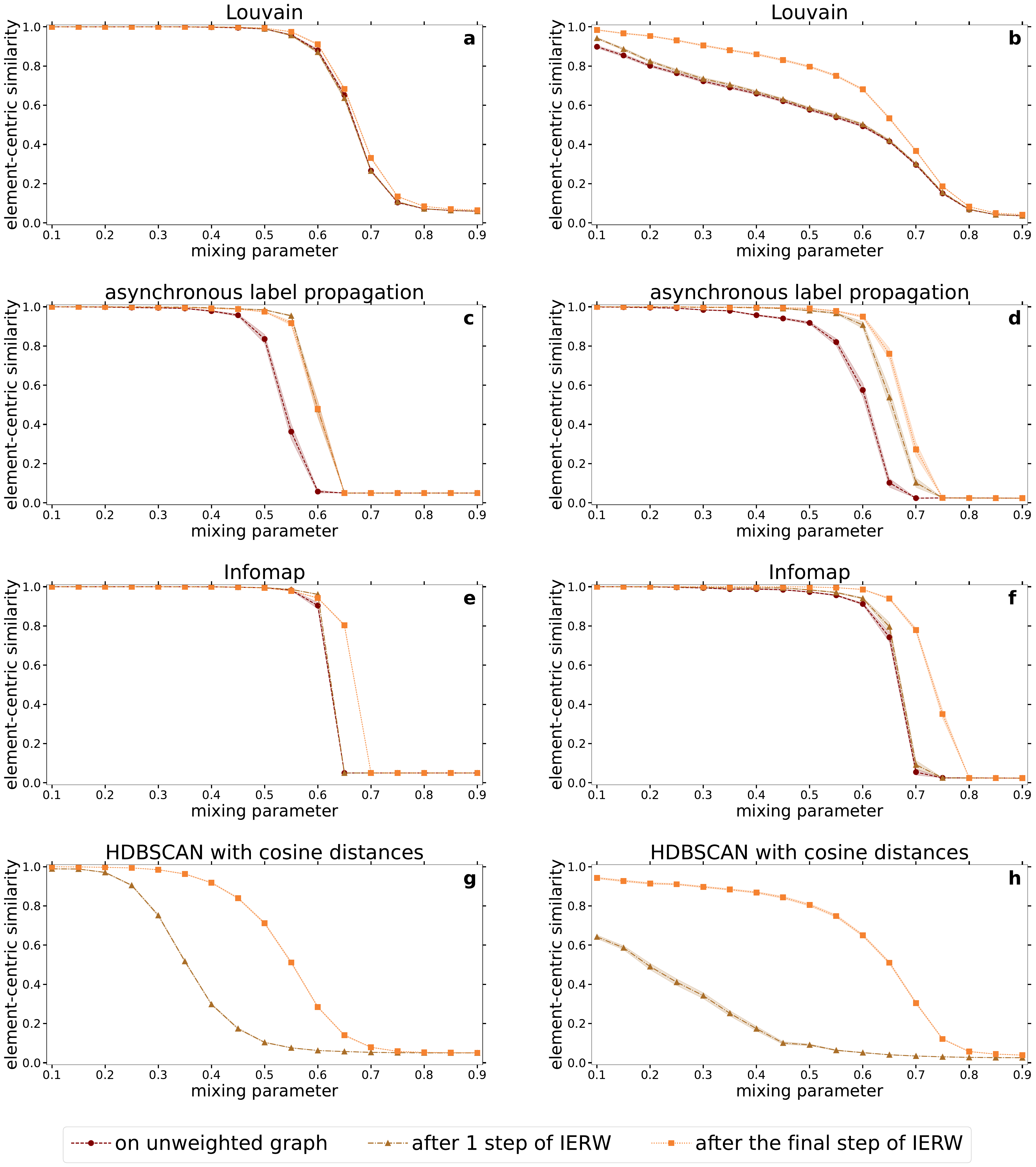}
    \caption{ {\bf Performance of standard clustering methods on the weighted networks and the embeddings derived by IERW using node2vec with exponentialized link weights.} Each row of panels corresponds to a different community detection method. The left column refers to networks generated by the PP model and the right one to networks generated by the LFR benchmark. We performed the community detection with all the methods only once for each network. Each displayed data point corresponds to a result averaged over $100$ networks, and the error bars depict the standard error of the mean.}
    \label{fig:tradCommDetHelpedWithExpNode2vec}
\end{figure}

As it is shown in Fig.~\ref{fig:flowchart}, our IERW process can aid community detection in two different ways: one may either apply standard data clustering techniques on the spatial node arrangements obtained from the embedding steps, or opt for community-finding methods developed for weighted networks, taking into account both the network topology and the geometric relations of the embedded nodes. In Fig.~\ref{fig:tradCommDetHelpedWithExpNode2vec}, we show examples for both options. On the one hand, we compare the performance of traditional network community-finding approaches on unweighted synthetic benchmark graphs to the results achieved when these methods are augmented by the link weights obtained from a single and multiple iterations of IERW using node2vec. As the network community detection methods, we employed Louvain~\cite{Louvain,Louvain_code} (Fig.~\ref{fig:tradCommDetHelpedWithExpNode2vec}a,b), asynchronous label propagation~\cite{alabprop,alabprop_code} (Fig.~\ref{fig:tradCommDetHelpedWithExpNode2vec}c,d) and Infomap~\cite{Infomap,Infomap_code} (Fig.~\ref{fig:tradCommDetHelpedWithExpNode2vec}e,f). In addition, we tested Hierarchical Density-Based Spatial Clustering of Applications with Noise (HDBSCAN)~\cite{HDBSCAN1,HDBSCAN2,HDBSCAN_code} on both the first and the iterated node embeddings (Fig.~\ref{fig:tradCommDetHelpedWithExpNode2vec}g,h), inputting only the cosine distance between all the possible node pairs in the embedding space. The left column of Fig.~\ref{fig:tradCommDetHelpedWithExpNode2vec} displays the ECS scores achieved for the PP test graphs of Figs.~\ref{fig:angSepRate_PP} and \ref{fig:commDetWithWeightThreshold}a--c, whereas the right column of Fig.~\ref{fig:tradCommDetHelpedWithExpNode2vec} refers to the LFR networks examined in Figs.~\ref{fig:angSepRate_LFR} and \ref{fig:commDetWithWeightThreshold}d--f. We repeated the experiments shown in Fig.~\ref{fig:tradCommDetHelpedWithExpNode2vec} using LE, ISO and TREXPIC embeddings too: the results, qualitatively very similar, are shown in Sect.~\ref{sect:tradCommDet_spectral_supp} of the Supplementary Information.

Regarding traditional network community detection methods, it is important to keep in mind that while Louvain, asynchronous label propagation and Infomap expect proximity-like link weights, the link weights $w_{ij}$ provided by IERW can be both distance-like (when using LE, ISO and TREXPIC) and proximity-like (in the case of node2vec). Hence, following a similar practice to the one suggested in Ref.~\cite{coalescentEmbedding}, in Sect.~\ref{sect:tradCommDet_spectral_supp} of the Supplementary Information we used a conversion formula 
\begin{equation}
    \tilde{w}_{ij}=\frac{1}{w_0+w_{ij}}
    \label{eq:distProxConvForTradMethods_main}
\end{equation}
on the link weights obtained from IERW with LE, ISO and TREXPIC before applying Louvain, asynchronous label propagation or Infomap, where $w_{0}>0$ is a tunable parameter. In general, by choosing a small $w_0$ we put more emphasis on the distances close to 0, in agreement with the expectation that the distances within communities eventually decrease over the iterations. Our analysis detailed in Sect.~\ref{sect:tradCommDet_spectral_supp} of the Supplementary Information shows that $w_0$ can affect the performance of the network community-finding methods when using IERW with LE, ISO and TREXPIC. Similarly, we also used a conversion formula 
\begin{equation}
    \tilde{w}_{ij}=w_0+w_{ij}
    \label{eq:proxProxConvForTradMethods}
\end{equation}
after applying IERW with node2vec, setting $w_0$ to $1.0$ in Fig.~\ref{fig:tradCommDetHelpedWithExpNode2vec}a--f, as we found that this shifting of all the proximity-like exponential link weights provided by IERW can improve the performance of all the examined traditional network community detection methods.

As it can be seen in Fig.~\ref{fig:tradCommDetHelpedWithExpNode2vec}, the node2vec-based IERW process can strongly improve the performance of standard clustering methods. We observed the largest improvement in the case of Louvain, when applied to LFR networks (Fig.~\ref{fig:tradCommDetHelpedWithExpNode2vec}b). It is well-known that community-finding methods based on modularity maximization (such as Louvain) may fail in detecting small communities~\cite{resolution_limit_PNAS}. Since the size distribution of the communities is relatively broad in the examined LFR networks, the ECS achieved on the original unweighted test graph (dark red curve) remains well below $1$ already at low $\mu$ values in Fig.~\ref{fig:tradCommDetHelpedWithExpNode2vec}b, indicating that Louvain in itself cannot fully uncover the planted community structure. The performance after only a single embedding (light brown curve) is similar to what is achieved in the unweighted case. However, when switching to the weighted networks provided by the complete process of IERW (orange curve), the performance greatly improves. Note that in the similar measurements performed with LE, TREXPIC and ISO in Figs.~\ref{fig:tradCommDetHelpedWithLE_optimalConsts}--\ref{fig:tradCommDetHelpedWithExpISO_optimalConsts} of the Supplementary Information, IERW seems to actually eliminate the resolution limit of modularity optimization, increasing the ECS of Louvain to $1$ in a wide range of the mixing parameter.


In the case of Louvain applied to PP networks (Fig.~\ref{fig:tradCommDetHelpedWithExpNode2vec}a) and Infomap (Fig.~\ref{fig:tradCommDetHelpedWithExpNode2vec}e,f), the results on the original, unweighted input graphs are already of very high quality. However, a slight increase can still be observed here when switching to the networks weighted by IERW with node2vec. In the case of asynchronous label propagation (Fig.~\ref{fig:tradCommDetHelpedWithExpNode2vec}c,d), the performance of a single embedding is similar to that of the iterated embedding, both being significantly better compared to the unweighted case. Finally, when applying HDBSCAN to the spatial node arrangements created by node2vec (Fig.~\ref{fig:tradCommDetHelpedWithExpNode2vec}g,h), although the performance after a single embedding is modest, the iteration of the embedding yields major improvements for both the PP and the LFR graphs.


\section*{Discussion}\label{sect:discussion}

We have shown that graph embeddings facilitate the identification of communities, by providing distance- or proximity-based weights to the links of the input graph, which makes its community structure topologically more pronounced and more easily detectable. At the same time, embedding graphs with progressively stronger community structures makes communities more apparent also in the embedding space, where they appear as clouds of points that become more and more compact and separated from each other. These observations inspired our IERW framework, which realizes a simple iterative procedure to ease community detection, where the network is repeatedly embedded and reweighted based on the geometric distance between the endpoints of the links. For embedding methods such as node2vec, where a larger link weight is interpreted as the indicator of a stronger and closer connection, as we keep iterating, intra-community link weights get larger and inter-community link weights get smaller. For the other embedding methods studied in the present paper (where the link weights are assumed to be distance-like), IERW acts in the opposite manner, increasing the weight of inter-community links and decreasing the weight of intra-community links over the iterations. 
Both cases suggest a simple way to find the clusters: removing inter-community links via weight thresholding. Such an approach, albeit elementary, is competitive with state-of-the-art community detection techniques. 

We stress that we only reweight the links of the original graph. If we assigned a weighted link to each pair of nodes, whether they are connected or not, the identification of the communities may 
become easier but at the cost of having a procedure with at least quadratic complexity in the number of nodes $N$. By focusing on the actual links of the input network, instead, the reweighting procedure has linear complexity in the number of links, which is much lower than $N^2$ on sparse networks. The ultimate complexity of the repeated embedding and weighting steps 
is determined by the running time of the chosen embedding algorithm. In the case of node2vec, for instance, the complexity of IERW would be $\pazocal{O}\left(E+N\cdot d\cdot\omega^2\right)$ for a network of $N$ nodes and $E$ edges when using a $d$-dimensional embedding space and $\omega$ window length. Here the results are fairly stable as a function of $d$, so one can pick a fixed value (we used $d=64$ in our experiments). For the other examined embeddings, there is a much stronger dependence on the number of embedding dimensions, and identifying a good range may be costly (see Methods).

Our method could be used as a pre-processing step in a community detection pipeline. We find that a single iteration of IERW can already produce a weighted network having stronger communities than the original graph. Applied after IERW, standard community detection techniques generally deliver better results than when they operate on the initial graph. Interestingly, our reweighting strategy provides a way to mitigate the effect of 
the resolution limit of modularity maximization, significantly improving the performance of such methods on realistic benchmarks.

Finally, we would like to stress that techniques like IERW could help facilitating other tasks, besides community detection. It would be interesting, for instance, to check whether link prediction also becomes easier on the weighted graphs and/or embeddings built by IERW or similar procedures, bearing in mind that different tasks may need different weighting rules and the application of different geometric measures. 

\section*{Methods}
\label{sect:methods}


\subsection*{Node embedding with Laplacian Eigenmaps}
\label{subsect:LE_main}

Based on the eigendecomposition of the Laplacian matrix of a neighborhood graph made from the original data set, the Laplacian Eigenmaps (LE) approach was first devised in Ref.~\cite{LE} for mapping data points supplied in a high-dimensional space onto a lower dimensional one. When applied to a weighted network, in the first step the assumed distance-like input weights $w_{ij}$ are converted to proximity-like weights using the exponential formula $w_{ij}'(w_{ij})=\exp(-w_{ij}^2/t)$, where, following the implementation created for Ref.~\cite{coalescentEmbedding}, we set the parameter $t$ to be equal to the square of the mean of the distance-like weights. 
Then, from the corresponding adjacency matrix $\mathbfit{A}$ and the diagonal matrix $\boldsymbol{\pazocal{D}}$ with $\pazocal{D}_{ii}=\sum_j A_{ij}$, we can obtain the Laplacian matrix as $\mathbfit{L}=\boldsymbol{\pazocal{D}}-\mathbfit{A}$. The eigenvectors $f_1,f_2,\dots,f_d$ satisfying the generalized eigenvector problem $\mathbfit{L}\cdot\underline{f}_{\ell}=\lambda_{\ell}\cdot\boldsymbol{\pazocal{D}}\cdot\underline{f}_{\ell}$ with the smallest non-zero eigenvalues ${\lambda_1\leq\lambda_2\leq...\leq\lambda_d}$ naturally define an embedding in the $d$-dimensional Euclidean space, where the $\ell^{\rm th}$ coordinate of the $i^{\rm th}$ node is given by the $i^{\rm th}$ component of $f_{\ell}$, making strongly connected nodes being as close to each other as possible.

The computational complexity of LE is $\pazocal{O}\left((d+1)\cdot N^2\right)$, where the dominant contribution comes from the eigendecomposition of the $N\times N$-sized graph Laplacian. A fully detailed algorithmic description of LE is provided in Sect.~\ref{sect:embAlgorithms} of the Supplementary Information.


In IERW with LE, we defined the distance-like input weights based on the angular distance $\Delta\theta_{ij}$ in the previous embedding as $w_{ij}=1-\cos(\Delta\theta_{ij})$.

\subsection*{Node embedding with TRansformation of EXponential shortest Path lengths to hyperbolIC measures}
\label{subsect:TREXPIC_main}

The TRansformation of EXponential shortest Path lengths to hyperbolIC measures (TREXPIC) method~\cite{TREXPIC} embeds networks in a $d$-dimensional hyperbolic space, trying to express the topological node-node distances as hyperbolic distances. First, TREXPIC prepares a matrix $\mathbfit{X}$ of expected hyperbolic distances based on the shortest path lengths ${\rm SPL}_{ij}$ measured along the graph, using the exponential formula $X_{ij}=\exp(-t/{\rm SPL}_{ij})$. Here we set the parameter $t>0$ to the default value defined in Ref.~\cite{TREXPIC}, given by $t=\sqrt{\ln(1.0/0.9999)\cdot\ln(1.0/0.1)}\cdot\mathrm{SPL}_{\mathrm{max}}$ with $\mathrm{SPL}_{\rm max}$ being the maximal shortest path length found in the network. 
The distance matrix $\mathbfit{X}$ is then converted into the matrix $\boldsymbol{\pazocal{L}}$ of expected pairwise Lorentz products, using the formula $\pazocal{L}_{ij}=\cosh(\zeta\cdot X_{ij})$, where we set $\zeta$ simply to $1$, and thus, the curvature of the hyperbolic space $K=-\zeta^2$ to $-1$. 
Finally, the matrix $\boldsymbol{\pazocal{L}}$ is subjected to singular value decomposition (formulated as $\boldsymbol{\pazocal{L}}=\mathbfit{U}\cdot\mathbfit{\Sigma}\cdot\mathbfit{V}^{\mathrm{T}}$): the length of the node position vectors is calculated from the largest singular value $\sigma_1\equiv\Sigma_{11}$ and the corresponding singular vector $\underline{u}_1$ (given by the first column of the matrix $\mathbfit{U}$), while the direction vectors of the embedded nodes are calculated from the next $d$ singular values ($\sigma_2\geq\sigma_3\geq...\geq\sigma_{d+1}$) and the corresponding singular vectors ($\underline{u}_2,\,\underline{u}_3,\,...,\,\underline{u}_{d+1}$).

The computational complexity of TREXPIC for ${d+1<\ln(N)}$ is dominated by the calculation of the $N\times N$-sized shortest path length matrix, yielding $\pazocal{O}\left(\ln(N)\cdot N^2\right)$, while the computational complexity of the truncated singular value decomposition is $\pazocal{O}\left((d+1)\cdot N^2\right)$. A fully detailed description of the TREXPIC approach is presented in Sect.~\ref{sect:embAlgorithms} of the Supplementary Information.

Similarly to the case of LE, in IERW with TREXPIC we defined the distance-like input weights as $ w_{ij}=1-\cos(\Delta\theta_{ij})$ based on the previous embedding iteration.

\subsection*{Node embedding with Isomap}
\label{subsect:ISO_main}

Similarly to LE, the Isomap (ISO) method was originally proposed~\cite{Isomap} for finding a lower-dimensional representation of a high-dimensional data set using a nearest neighbor graph. Aiming at a mapping between the topological node-node distances and the Euclidean distances in the embedding, a matrix $\mathbfit{I}$ of expected pairwise inner products is calculated from the shortest path length (SPL) matrix of the graph to be embedded, placing the center of mass of the embedded graph at the origin. In the present paper, we followed the implementation applied in Ref.~\cite{coalescentEmbedding}, which performs not the eigendecomposition but the singular value decomposition of the matrix $\mathbfit{I}$. 
This singular value decomposition (formulated as $\mathbfit{I}=\mathbfit{U}\cdot\mathbfit{\Sigma}\cdot\mathbfit{V}^{\mathrm{T}}$) provides the node coordinates in the $d$-dimensional Euclidean space: by taking the $d$ largest singular values ${\sigma_1\geq\sigma_2\geq...\geq\sigma_d}$ and the corresponding singular vectors $\underline{u}_1,\,\underline{u}_2,\,...,\,\underline{u}_d$, the $\ell^{\rm th}$ component of the position vector of the $i^{\rm th}$ network node is defined as $\underline{y}_i(\ell)=\sqrt{\sigma_{\ell}}\cdot\underline{u}_{\ell}(i)$. 

To improve the performance of IERW, we introduced an alternative version of ISO that is built on exponentialized shortest path lengths, similarly to TREXPIC. Here, the original formula $D_{ij}=\mathrm{SPL}_{ij}$ of the expected pairwise Euclidean distances is replaced by $D_{ij}=\exp(-t/{\rm SPL}_{ij})$, where $t>0$ is a tunable parameter. 
We used the same setting for this $t$ parameter as in the default case of TREXPIC, namely ${t=\sqrt{\ln(1.0/0.9999)\cdot\ln(1.0/0.1)}\cdot\mathrm{SPL}_{\mathrm{max}}}$, where $\mathrm{SPL}_{\mathrm{max}}$ is the largest shortest path length of the examined network. The beneficial effect of exponentialization in ISO is demonstrated in Sect.~\ref{sect:expVSnotExp} of the Supplementary Information.

The computational complexity of ISO for ${d<\ln(N)}$  dominated by the calculation of the $N\times N$ shortest path length matrix, yielding $\pazocal{O}\left(\ln(N)\cdot N^2\right)$, while the computational complexity of the truncated singular value decomposition is $\pazocal{O}\left(d\cdot N^2\right)$. A fully detailed description of ISO embeddings is given in Sect.~\ref{sect:embAlgorithms} of the Supplementary Information.

In complete analogy with LE and TREXPIC, in IERW with both versions of ISO we defined the link weights of the network based on the previous embedding iteration simply as $w_{ij}=1-\cos(\Delta\theta_{ij})$.

\subsection*{Node embedding with node2vec}
\label{subsect:node2vec_main}

The node2vec method~\cite{node2vec} provides Euclidean node embeddings based on random walks in the network. The central idea is to use the sequences of the visited nodes as textual input for the word2vec~\cite{word2vec} method, originally designed to embed words from a large text corpus into a vector space. 
In the present paper, we followed the parameter setting proposed in Ref.~\cite{commDetInEmbs_Santo} by setting the number of walks started from each node to $80$, the length of the random walk to $10$ and the length of the considered context windows in word2vec to $\omega=10$. The parameters $p$ and $q$, controlling the locality and the depth of the random walks were set to the default value of $p=q=1$.  

The computational complexity of creating a $d$-dimensional embedding for a network of $N$ nodes and $E$ edges with node2vec is $\pazocal{O}\left(E+N\cdot d\cdot\omega^2\right)$. Note that since node2vec operates with random walks, it is a stochastic embedding method. Nonetheless, as we performed all of our measurements for multiple network samples anyway, we ran IERW with node2vec only once for each network. A more detailed description of the node2vec method is given in Sect.~\ref{sect:embAlgorithms} of the Supplementary Information.

When provided with a weighted input network, the random walk transition probabilities are modified in node2vec according to the link weights, where a higher link weight is accompanied by a higher transition probability. According to that, opposite to the previous embedding methods, node2vec expects proximity-like link weights instead of distance-like weights. To utilize the beneficial effects of exponentialization in IERW also with node2vec, here we defined exponential link weights based on the previous embedding iteration as $w_{ij}=\exp\left[t\cdot (\cos(\Delta\theta_{ij})-1)\right]$, where the parameter $t$ was set to $t=10\cdot\bar{\kappa}/\hat{\kappa}$ with $\bar{\kappa}$ denoting the average and $\hat{\kappa}$ the mode of the node degrees, respectively. 
The advantage of the exponentialization over the application of a simple proximity-like link weight formula given by $w_{ij}=\cos(\Delta\theta_{ij})+1$ is demonstrated in Sect.~\ref{sect:expVSnotExp} of the Supplementary Information.

\subsection*{Choosing the number of embedding dimensions}
\label{subsect:chooseEmbDim_main}

When using IERW with node2vec, we followed one of the standard choices in the literature and simply set the number of embedding dimensions $d$ always to $64$. However, as it is demonstrated in Sect.~\ref{sect:embParams} of the Supplementary Information, the performance of LE, ISO and TREXPIC shows a relatively strong dependence on the setting of $d$, and in the case of these matrix decomposition methods, it seems that the best choice is a $d$ close to the number of communities in the examined network. Therefore, before applying IERW with LE, ISO or TREXPIC, we estimated the number of planted communities $C$ based on the number of non-zero eigenvalues below the largest eigengap of the normalized Laplacian matrix of the given network, and using this estimation, we set the number of embedding dimensions to $d=C-1$, which fits the expectation that e.g. a two-dimensional pattern (namely a triangle) is needed in general to describe all the pairwise relations between three communities. The algorithmic details of choosing the number of embedding dimensions for LE, ISO and TREXPIC are provided in Sect.~\ref{sect:embParams} of the Supplementary Information.

\subsection*{Extraction of communities with weight thresholding}
\label{subsect:commDetAlgorithmWithWeightThreshold}

As described in Results, for demonstration purposes we implemented a really simple community detection method that performs a weight thresholding on the weighted networks obtained from the IERW process. Namely, we aimed at splitting a network into groups of densely connected nodes through the following steps:
\begin{enumerate}
    \item Sort the weights of the $E$ number of links of the embedded network in increasing order. 
    \item Remove the $\lfloor0.05\cdot E\rfloor$ lowest and the largest links from the ordered list to ensure the removal of at least $5\%$ but at most $95\%$ of the links during the weight thresholding.
    \item Find the largest gap between the consecutive link weights in the ordered list and set the weight threshold to the average of the two weight values on the sides of the largest gap.
    \item Prune the examined graph.
    \begin{enumerate}
        \item When dealing with distance-like link weights that are smaller for stronger connections (i.e., when using LE, TREXPIC or ISO), remove the links having weights larger than the threshold.
        \item When dealing with proximity-like link weights that are larger for stronger connections (i.e., when using node2vec), remove the links having weights smaller than the threshold.
    \end{enumerate}
    \item Identify each of the connected components in the pruned graph as a community.
\end{enumerate}

\subsection*{Generating synthetic networks with communities using the planted partition model}
\label{subsect:plantedPart}

The planted partition (PP) model~\cite{plantedPartitionModel} is a special case of the stochastic block model (SBM)~\cite{simplestSBMarticle}, where there are only two values for the link probability: $p_{\mathrm{in}}$, for pairs of nodes in the same community/block and $p_{\mathrm{out}}$ for pairs of nodes in different communities/blocks. To generate networks with the PP model, we used the Python function ‘planted\_partition\_graph’ available in the ‘NetworkX’ package.

The input parameters of the model are the total number $N$ of nodes, the number $m$ of nodes in each community and the expected average degree $\bar{\kappa}$. In the above-presented measurements, following the settings in Ref.~\cite{commDetInEmbs_Santo}, we used $N=1000$, $m=50$ (yielding $C=N/m=20$ communities) and $\bar{\kappa}=20$. 
To obtain community structures of different strengths, we tuned the mixing parameter ${\mu\in[0,1]}$, which we define as the fraction between the expected number of neighbors of a randomly chosen node outside of its community and the expected total number of neighbors, i.e. as $\mu=\bar{\kappa}_{\mathrm{out}}/\bar{\kappa}$. Given the mixing parameter $\mu$ and the expected average degree $\bar{\kappa}$, we calculated the expected number of inter-cluster edges of each node as $\bar{\kappa}_{\mathrm{out}}=\mu\cdot\bar{\kappa}$, and the expected number of intra-cluster edges of each node as $\bar{\kappa}_{\mathrm{in}}=\bar{\kappa}-\bar{\kappa}_{\mathrm{out}}$. Then, we derived the desired connection probabilities $p_{\mathrm{out}}$ and $p_{\mathrm{in}}$ from the formulas ${\bar{\kappa}_{\mathrm{out}}=p_{\mathrm{out}}\cdot(C-1)\cdot m}$ and $\bar{\kappa}_{\mathrm{in}}=p_{\mathrm{in}}\cdot(m-1)$. Self-loops are not included in the applied implementation, meaning that the number of possible neighbors of a node within its own block is $m-1$ instead of $m$. In our measurements, we used the settings $\mu=0.1,\,0.15,\,0.2,\,...,\,0.85,\,0.9$, where smaller values correspond to more easily detectable community structures.


\subsection*{Generating synthetic networks with communities using the Lancichinetti--Fortunato--Radicchi benchmark}
\label{subsect:LFR}

The Lancichinetti--Fortunato--Radicchi (LFR) benchmark~\cite{LFRbenchmark} generates graphs with power-law distributions of the node degrees and the community sizes, enabling the emergence of heterogeneity in these two quantities. The input parameters of the model are the total number $N$ of nodes, the expected average degree $\bar{\kappa}$, the allowed largest degree $\kappa_{\mathrm{max}}$, the exponent $\gamma$ of the tail of the degree distribution ($\pazocal{P}(\kappa)\sim \kappa^{-\gamma}$), the allowed smallest and largest community sizes $m_{\mathrm{min}}$ and $m_{\mathrm{max}}$, the exponent $\beta$ of the tail of the community size distribution ($\pazocal{P}(m)\sim m^{-\beta}$), and the mixing parameter ${\mu\in[0,1]}$, having the same definition that we used in the case of the PP model, meaning that each node is expected to share a fraction of $1-\mu$ of its links with the other nodes of its own community and the remaining fraction $\mu$  with the nodes of the other communities. 
We examined LFR networks with non-overlapping clusters that we generated with the C++ code downloaded from \url{https://www.santofortunato.net/resources}. In the above measurements, following the settings in Ref.~\cite{commDetInEmbs_Santo}, we used $N=1000$, $\bar{\kappa}=20$, $\kappa_{\mathrm{max}}=50$, $\gamma=2.0$, $m_{\mathrm{min}}=10$, $m_{\mathrm{max}}=100$ and $\beta=3.0$, tuning the mixing parameter between $\mu=0.1$ (yielding easily detectable community structures with links falling mostly within communities) and $\mu=0.9$ (where most of the links connect nodes of different communities). 

\section*{Code availability}
\begin{sloppypar}
The code used for performing the iterative embedding is available at \url{https://github.com/BianKov/iterEmb}.
\end{sloppypar}

\section*{Data availability}
All data generated during the current study are available from the corresponding author upon request.



\section*{Acknowledgements}

This project has received funding from the European Union’s Horizon 2020 research and innovation programme under grant agreement no. 101021607, 
the European Union project RRF-2.3.1-21-2022-00004 within the framework of the Artificial Intelligence National Laboratory
and was partially supported by the National Research, Development and Innovation Office under grant no. K128780.
We acknowledge the support of the AccelNet-MultiNet program, a project of the National Science Foundation (Award \#1927425 and \#1927418),
of the Army Research Office under Contract No. W911NF-21-1-0194,
and of the National Institutes of Health under awards U01AG072177 and U19AG074879.

\section*{Author contributions}
S.F. and G.P. developed the concept of the study, B.K. worked out the details of the iterative embedding algorithms, performed the measurements and prepared the figures, B.K., G.P., S.F. and S.K. analysed and interpreted the results, S.K. optimized the implementation of the algorithms, B.K., G.P., S.F. and S.K. wrote the paper. All authors reviewed the manuscript.

\section*{Ethics declarations}
\subsection*{Competing interests}
The authors declare no competing interests.

\clearpage

\begin{flushleft} 
{\huge \bfseries{SUPPLEMENTARY INFORMATION}}
\end{flushleft}

\setcounter{section}{0}
\setcounter{equation}{0}
\setcounter{figure}{0}
\setcounter{table}{0}
\renewcommand{\thesection}{S\arabic{section}}
\renewcommand{\thefigure}{S\arabic{figure}}
\renewcommand{\thetable}{S\arabic{table}}
\renewcommand{\theequation}{S\arabic{equation}}

\renewcommand{\thesection}{S1}
\section{The applied embedding algorithms in detail}
\label{sect:embAlgorithms}
\setcounter{subsection}{0}

This section provides the exact definition of the four node embedding algorithms that we examined. Section~\ref{subsect:LE} describes Laplacian Eigenmaps (LE), Sect.~\ref{subsect:TREXPIC} deals with the hyperbolic method named TRansformation of EXponential shortest Path lengths to hyperbolIC measures (TREXPIC), Sect.~\ref{subsect:ISO} presents Isomap (ISO) and its exponentialized version, and Sect.~\ref{subsect:node2vec} details node2vec. At the end of each subsection, the weight formula applied with the given embedding algorithm in IERW is also explained. Note that while LE, TREXPIC and ISO expect distance-like link weights (where higher values refer to weaker connection),
node2vec expects proximity-like link weights (where a higher value
indicates the higher strength or relevance of the given connection).

\subsection{Node embedding with Laplacian Eigenmaps}
\label{subsect:LE}

The Laplacian Eigenmaps (LE) method was originally developed in Ref.~\cite{LE} for mapping data points given in a high-dimensional space to a lower-dimensional space based on the eigendecomposition of the Laplacian matrix of a nearest neighbor graph created from the original data set. Considering a real complex network as a graph constructed in a high-dimensional space following the distance relations between the data points, the steps carried out in the LE algorithm after creating the neighborhood graph can be applied for obtaining a spatial representation of the network topology~\cite{LE_hypEmbedding_firstPSObased,LE_hypEmbedding_PSObasedRwithLLoptim,coalescentEmbedding,Mercator_H2model,dMercator}, yielding the following node embedding algorithm:
\begin{enumerate}
    \item If link weights $w_{ij}\geq0$ are given, these are interpreted by the algorithm as the distance between the connected nodes $i,\,j$ in the latent high-dimensional space, meaning that a larger link weight is read as a weaker connection between two nodes. The inputted distance-like weights are converted to proximity-like weights $w_{ij}'\in(0,1]$ using the exponential formula
    \begin{equation}
        w_{ij}'(w_{ij})=e^{-\frac{w_{ij}^2}{t}},
        \label{eq:LEdistProxConv}
    \end{equation}
    where, following the implementation created for Ref.~\cite{coalescentEmbedding}, we calculated the scaling factor $t$ as the square of the mean of the distance-like weights. 
    When using our IERW process with the LE method, we re-set the $t$ parameter in each iteration according to the current link weights.
    \item The adjacency matrix $\mathbfit{A}$ is created. 
        \begin{enumerate}
           \item For unweighted graphs, $A_{ij}\equiv A_{ji}=0$ if node $i$ is not connected to node $j$ and otherwise $A_{ij}\equiv A_{ji}=1$.
           \item For weighted graphs, $A_{ij}\equiv A_{ji}=0$ if and only if nodes $i$ and $j$ are not connected to each other, and $A_{ij}\equiv A_{ji}=w_{ij}'$ otherwise. Note that links with $0$ distance-like weight yield non-zero elements in the adjacency matrix $\mathbfit{A}$ since $w_{ij}'(0)=1$.
       \end{enumerate}
    As self-loops are not considered, $A_{ii}=0$ for any node index $i$ in both cases.
    \item The Laplacian matrix $\mathbfit{L}=\boldsymbol{\pazocal{D}}-\mathbfit{A}$ is calculated, where $\boldsymbol{\pazocal{D}}$ is the diagonal matrix of elements $\pazocal{D}_{ii}=\sum_j A_{ij}$.
    \item Given the generalized eigenvector problem $\mathbfit{L}\cdot\underline{f}_{\ell}=\lambda_{\ell}\cdot\boldsymbol{\pazocal{D}}\cdot\underline{f}_{\ell}$, the eigenvectors $\underline{f}_1,\,\underline{f}_2,\,...,\,\underline{f}_d$ corresponding to the eigenvalues ${\lambda_1\leq\lambda_2\leq...\leq\lambda_d}$ are computed. The first eigenvalue $\lambda_0$ in the increasing order of eigenvalues is always $0$, and the corresponding eigenvector $\underline{f}_0$ is not used in the embedding.
    \item Each eigenvector is used as a list of Cartesian coordinates of all the embedded network nodes along a given axis of the $d$-dimensional Euclidean space: the $\ell^{\rm th}$ ($\ell=1,\,2,\,...,\,d$) Cartesian coordinate in the position vector $\underline{y}_i$ of the graph's $i^{\rm th}$ ($i=1,\,2,\,...,\,N$) node is set to the $i^{\rm th}$ element of the eigenvector $\underline{f}_{\ell}$, i.e. $\underline{y}_i(\ell)=\underline{f}_{\ell}(i)$.
\end{enumerate}
For a network of $N$ number of nodes, the computational complexity of embedding a network in the $d$-dimensional Euclidean space using the above algorithm is $\pazocal{O}\left((d+1)\cdot N^2\right)$, where the dominant step is the eigendecomposition of the graph Laplacian.

LE focuses on preserving local information, namely the neighborhood relations by putting the (strongly) connected nodes as close to each other in the embedding space as possible: it assigns to every node $i$ a position vector $\underline{y}_i$ in such a way that a weighted sum $\sum_{i=1}^N\sum_{j=1}^N A_{ij}\cdot\|\underline{y}_i-\underline{y}_j\|^2$ of the squared Euclidean distances between the connected node pairs is minimized while preventing the collapse of all the network nodes into a single point~\cite{LE,Mercator_H2model}. In our iterative embedding process, we made LE placing nodes with small angular distance $\Delta\theta_{ij}$ closer to each other with respect to the Euclidean distance in the subsequent iteration by using the formula
\begin{equation}
    w_{ij}=1-\cos(\Delta\theta_{ij}),
    \label{eq:weightForLE}
\end{equation} 
for creating distance-like link weights. Since placing nodes close to each other in the Euclidean space inherently makes their angular separation small, the iteration of LE using the weight formula of Eq.~(\ref{eq:weightForLE}) can increase the separation between the angularly arranged communities.

\subsection{Node embedding with TRansformation of EXponential shortest Path lengths to hyperbolIC measures}
\label{subsect:TREXPIC}

The method named TRansformation of EXponential shortest Path lengths to hyperbolIC measures (TREXPIC) was introduced in Ref.~\cite{TREXPIC} for embedding networks in the hyperbolic space, even considering the possible directedness of the links. The hyperbolic space is commonly claimed to be a particularly good candidate for hosting graphs that have an underlying hierarchical orderliness or tree-like structure. The presence of hierarchy is usually understood broadly, as a general consequence of the presence of some kind of heterogeneity of the network nodes, which implies the possibility for classifying the nodes into groups that are interconnected through a containment hierarchy~\cite{hyperGeomBasics}, or yields a natural ranking of the network nodes where the status of each node is determined by the given heterogeneous topological property~\cite{hypGeom_hierarchy_connection} (e.g. the node degree, which can be treated as a simple measure of importance, and thus, as a proxy for a node's status in the hierarchy of importance~\cite{degreeBasedHierarchy}).

TREXPIC places the network nodes in the so-called native representation~\cite{hyperGeomBasics} of the hyperbolic space that visualizes the negatively curved hyperbolic space in the flat Euclidean space simply as a ball of infinite radius that we call the native ball or, in the two-dimensional case, the native disk. The aim of TREXPIC is the reconstruction of a matrix $\mathbfit{X}$ describing node-node distances measured along the links of a network in the form of the matrix of pairwise hyperbolic distances between the embedded network nodes. The hyperbolic distance $x_{\underline{y},\underline{z}}$ between two points given by the Cartesian coordinate vectors ${\underline{y}=\left[\underline{y}(1),\underline{y}(2),...,\underline{y}(d)\right]}$ and $\underline{z}=\left[\underline{z}(1),\underline{z}(2),...,\underline{z}(d)\right]$ in the $d$-dimensional native ball can be calculated from the hyperbolic law of cosines written as
\begin{equation}
    \mathrm{cosh}\left(\zeta x_{\underline{y},\underline{z}}\right)=\mathrm{cosh}(\zeta r_y)\,\mathrm{cosh}(\zeta r_z)-\mathrm{sinh}(\zeta r_y)\,\mathrm{sinh}(\zeta r_z)\,\mathrm{cos}(\Delta\theta_{y,z}),
    \label{eq:hypLawOfCosines}
\end{equation}
where $\zeta\in\mathbb{R}^{+}$ is connected to the curvature $K<0$ of the hyperbolic space as $\zeta=\sqrt{-K}$, $\Delta\theta_{y,z}=\mathrm{acos}\left(\frac{\underline{y}\cdot\underline{z}}{\|\underline{y}\|\,\|\underline{z}\|}\right)=\mathrm{acos}\left(\frac{\sum_{\ell=1}^d \underline{y}(\ell) \underline{z}(\ell)}{r_y r_z}\right)$ is the angular distance between the given two points, while $r_y\equiv\|\underline{y}\|=\sqrt{\sum_{\ell=1}^d \underline{y}(\ell)^2}$ and $r_z\equiv\|\underline{z}\|=\sqrt{\sum_{\ell=1}^d \underline{z}(\ell)^2}$ denote the radial coordinate of the examined two points. According to Eq.~(\ref{eq:hypLawOfCosines}), $r_y=0$ yields $x_{\underline{y},\underline{z}}=r_z$, and for $r_z=0$ simply $x_{\underline{y},\underline{z}}=r_y$, meaning that in the native representation, the hyperbolic distance measured from the origin $\underline{o}=[0,0,...,0]$ (which is the center of the native ball) is equal to the corresponding Euclidean distance, i.e. radial coordinate.

In addition to the native representation, TREXPIC also builds on the hyperboloid representation of the hyperbolic space in an intermediate step, just like a previous hyperbolic embedding method called hydra (hyperbolic distance recovery and approximation)~\cite{hydra}. The hyperboloid model represents the $d$-dimensional hyperbolic space in the $(d+1)$-dimensional Euclidean space as the upper sheet of a two-sheet hyperboloid. Here, the hyperbolic distance $x_{\underline{y},\underline{z}}$ between two points given by the Cartesian coordinate vectors $\underline{y}=\left[\underline{y}(1),\underline{y}(2),...,\underline{y}(d+1)\right]$ and $\underline{z}=\left[\underline{z}(1),\underline{z}(2),...,\underline{z}(d+1)\right]$ can be calculated as
\begin{equation}
    x_{\underline{y},\underline{z}}=\mathrm{acosh}\left(\underline{y}\circ\underline{z}\right)/\zeta,
    \label{eq:hypDist_hyperboloid}
\end{equation}
where $\underline{y}\circ\underline{z}$ is the Lorentz product
\begin{equation}
    \underline{y}\circ\underline{z} = \underline{y}(1)\underline{z}(1)-\left(\underline{y}(2)\underline{z}(2)+\underline{y}(3)\underline{z}(3)+...+\underline{y}(d+1)\underline{z}(d+1)\right)
    \label{eq:LorentzProductDefinition}
\end{equation}
between the two position vectors. The first coordinate $\underline{y}(1)$ of a $(d+1)$-dimensional position vector $\underline{y}$ given in the hyperboloid representation is measured along the hyperboloid's axis of rotation, and is always positive in the case of the upper sheet. It can be connected based on the formula $x_{\underline{y},\underline{y}}\equiv0$ --- that is equivalent to $\underline{y}\circ\underline{y}\equiv1$, as it can be seen from Eq.~(\ref{eq:hypDist_hyperboloid}) --- to the length of the $d$-dimensional vector formed by the coordinates from the second to the $(d+1)$th one as $\sqrt{\sum_{\ell=2}^{d+1}\underline{y}(\ell)^2}=\sqrt{\underline{y}(1)^2-1}$. Besides, the first coordinate $\underline{y}(1)$ can be used for computing the hyperbolic distance between any position given by $\underline{y}\in\mathbb{R}^{d+1}$ and the origin of the hyperbolic space that is given in the hyperboloid representation by the position vector $\underline{o}=[1,0,0,...,0]$: according to Eqs.~(\ref{eq:hypDist_hyperboloid}) and (\ref{eq:LorentzProductDefinition}), $x_{\underline{y},\underline{o}}=\mathrm{acosh}\left(\underline{y}\circ\underline{o}\right)/\zeta=\mathrm{acosh}\left(\underline{y}(1)\cdot 1-(\underline{y}(2)\cdot 0+\underline{y}(3)\cdot 0+...+\underline{y}(d+1)\cdot 0)\right)/\zeta=\mathrm{acosh}\left(\underline{y}(1)\right)/\zeta$.

The detailed steps of the algorithm of TREXPIC for embedding an undirected network in the $d$-dimensional hyperbolic space of curvature $K=-\zeta^2$ (where $\zeta\in\mathbb{R}^{+}$) are the following:
\begin{enumerate}
    \item Creation of the matrix $\mathbfit{X}$ of expected pairwise hyperbolic distances between the embedded network nodes according to the exponential formula
    \begin{equation}
        X_{ij}\equiv X_{ji}=e^{-\,\frac{t}{\mathrm{SPL}_{ij}}}\in[0,1),
        \label{eq:expSPLforTREXPIC}
    \end{equation}
    where the fast-changing, exponential mapping is intended to enable capturing the topological relations of the examined network more precisely. 
    \begin{enumerate}
        \item For unweighted graphs, $\mathrm{SPL}_{ij}$ is the possible smallest number of hops in which the $j$th network node can be reached from the $i^{\rm th}$ network node along the links.
        \item If link weights $w_{ij}\geq0$ are given, these are interpreted as the expected hyperbolic distance between the connected nodes $i,\,j$, meaning that a larger link weight is read as a weaker connection between two nodes, and two nodes connected with a link of $0$ weight --- which must not be confused with an unconnected node pair --- can access each other the easiest. In this case, $\mathrm{SPL}_{ij}$ is the possible smallest sum of link weights along the paths that connect the $i^{\rm th}$ network node to the $j$th one.
    \end{enumerate}
    In both cases, $\mathrm{SPL}_{ii}=0$, yielding $X_{ii}=0$ for any node index $i$. Following the default setting of the implementation of TREXPIC published for Ref.~\cite{TREXPIC}, we calculated the multiplying factor $t$ in Eq.~(\ref{eq:expSPLforTREXPIC}) from the largest shortest path length $\mathrm{SPL}_{\mathrm{max}}$ of the given network as $t=\sqrt{\ln(1.0/0.9999)\cdot\ln(1.0/0.1)}\cdot\mathrm{SPL}_{\mathrm{max}}$, corresponding to the geometric mean of two extreme settings given by $t_{\mathrm{small}}=\ln(1.0/0.9999)\cdot\mathrm{SPL}_{\mathrm{max}}$ (yielding a largest expected hyperbolic distance of $0.9999$) and $t_{\mathrm{large}}=\ln(1.0/0.1)\cdot\mathrm{SPL}_{\mathrm{max}}$ (yielding a largest expected hyperbolic distance of $0.1$). 
    When using our IERW process with the TREXPIC method, we re-set the $t$ parameter in each iteration according to the current $\mathrm{SPL}_{\mathrm{max}}$ yielded by the current link weights.
    \item Conversion of the matrix $\mathbfit{X}$ of expected hyperbolic distances to the matrix $\boldsymbol{\pazocal{L}}$ of the corresponding expected Lorentz products in the hyperboloid representation of the $d$-dimensional hyperbolic space of curvature $K=-\zeta^2$ according to Eq.~(\ref{eq:hypDist_hyperboloid}): ${\pazocal{L}_{ij}=\cosh(\zeta\cdot X_{ij})}$. Following the default setting in Ref.~\cite{TREXPIC}, we always set $\zeta$ to $1$.
    \item The derivation of the length and the direction of the position vector of each network node $i$ ($i=1,\,2,\,...,\,N$) in the $d$-dimensional native ball in accordance with the expected Lorentz products (and hereby also following the expected hyperbolic distances):
    \begin{enumerate}
        \item[i.] Find the largest $d+1$ number of singular values ${\sigma_1\geq\sigma_2\geq...\geq\sigma_{d+1}}$ and the corresponding singular vectors $\underline{u}_1,\,\underline{u}_2,\,...,\,\underline{u}_{d+1}$ of the Lorentz product matrix~$\boldsymbol{\pazocal{L}}$.
        \item[ii.] Use the largest singular value $\sigma_1$ and the corresponding singular vector $\underline{u}_1$ to calculate the $i^{\rm th}$ node's hyperbolic distance from the origin of the hyperbolic space: $x_{i,\underline{o}}=\mathrm{acosh}\left(\sqrt{\sigma_1}\cdot\underline{u}_1(i)\right)/\zeta$. Remark that the singular values are always non-negative, and the singular vector $\underline{u}_1$ corresponds to the leading eigenvector of the matrix $\boldsymbol{\pazocal{L}}\cdot \boldsymbol{\pazocal{L}}^{\mathrm{T}}$, which eigenvector --- according to the Perron--Frobenius theorem --- can be chosen to have only positive components. If some numerical errors yield $\sqrt{\sigma_1}\cdot\underline{u}_1(i)<1$, simply set $x_{i,\underline{o}}$ to $\mathrm{acosh}(1)/\zeta=0$.
        \item[iii.] Compute a $d$-dimensional direction vector for the $i^{\rm th}$ network node using the singular values and vectors from the second to the $(d+1)$th one: $\underline{e}_i=\left[\sqrt{\sigma_2}\cdot\underline{u}_2(i)\,,\,\sqrt{\sigma_3}\cdot\underline{u}_3(i)\,,\,...\,,\,\sqrt{\sigma_{d+1}}\cdot\underline{u}_{d+1}(i)\right]/\,n$, where $n$ is the norm $n=\sqrt{\left(\sqrt{\sigma_2}\cdot\underline{u}_2(i)\right)^2+\left(\sqrt{\sigma_3}\cdot\underline{u}_3(i)\right)^2+...+\left(\sqrt{\sigma_{d+1}}\cdot\underline{u}_{d+1}(i)\right)^2}$.
    \end{enumerate}
    Note that when embedding undirected networks, the Lorentz product matrix $\boldsymbol{\pazocal{L}}$ is symmetric, and thus, $\mathbfit{U}=\mathbfit{V}$ in the singular value decomposition $\boldsymbol{\pazocal{L}}=\mathbfit{U}\cdot\mathbfit{\Sigma}\cdot\mathbfit{V}^{\mathrm{T}}$, where $\mathbfit{\Sigma}$ is the diagonal matrix of singular values, and the columns of the matrixes $\mathbfit{U}$ and $\mathbfit{V}$ are the left and the right singular vectors of $\boldsymbol{\pazocal{L}}$, respectively. Besides, in the hyperboloid representation of the $(N-1)$-dimensional hyperbolic space, the $N\times N$-sized matrix $\boldsymbol{\pazocal{L}}$ of expected Lorentz products can be fully reconstructed according to its definition given by Eq.~(\ref{eq:LorentzProductDefinition}) --- i.e., as $\boldsymbol{\pazocal{L}}=\boldsymbol{\pazocal{Y}}\cdot\mathbfit{J}\cdot\boldsymbol{\pazocal{Y}}^{\mathrm{T}}$ with ${\mathbfit{J}=\mathrm{diag}(+1,-1,-1,...,-1)}$ of size $N\times N$ --- if the $N\times N$-sized node coordinate matrix $\boldsymbol{\pazocal{Y}}$ is defined based on the singular value decomposition $\boldsymbol{\pazocal{L}}=\mathbfit{U}\cdot\mathbfit{\Sigma}\cdot\mathbfit{U}^{\mathrm{T}}$ as ${\boldsymbol{\pazocal{Y}}=[\sqrt{\sigma_1}\cdot\underline{u}_1\,,\,\mathrm{i}\cdot\sqrt{\sigma_2}\cdot\underline{u}_2\,,\,\mathrm{i}\cdot\sqrt{\sigma_3}\cdot\underline{u}_3\,,\,...\,,\,\mathrm{i}\cdot\sqrt{\sigma_N}\cdot\underline{u}_N]}$ with $\mathrm{i}=\sqrt{-1}$ denoting the imaginary unit. The imaginary multiplying factors in the above formula of $\boldsymbol{\pazocal{Y}}$ do not raise any issues since the direction described by the coordinates from the second one is the same when all of these coordinates are purely imaginary as if they all were real numbers. When embedding in the native ball of $d<N-1$ number of dimensions, the directions that have smaller contribution in the Lorentz products --- i.e., where the singular value is smaller --- are neglected, while the hyperbolic distance from the origin of the hyperbolic space is preserved in any number of dimensions by setting it always according to the largest singular value $\sigma_1$ and the corresponding singular vector $\underline{u}_1$. 
    A similar approach is used by the hydra embedding algorithm~\cite{hydra} that also separates from each other a so-called "directional projection" and a "radial projection" of some $(d+1)$-dimensional vectors obtained from a decomposition of a matrix of expected Lorentz products.
    \item Calculation of the $i^{\rm th}$ ($i=1,\,2,\,...,\,N$) network node's $\ell^{\rm th}$ ($\ell=1,\,2,\,...,\,d$) Cartesian coordinate in the native representation of the $d$-dimensional hyperbolic space as $\underline{y}_i(\ell)=x_{i,\underline{o}}\cdot\underline{e}_i(\ell)$.
\end{enumerate}
The computational complexity of embedding a network of $N$ number of nodes in the $d$-dimensional hyperbolic space with the above algorithm is dominated at ${d+1<\ln(N)}$ by the calculation of all the shortest path lengths in the network, yielding $\pazocal{O}\left(\ln(N)\cdot N^2\right)$, while the computational complexity of the truncated SVD is $\pazocal{O}\left((d+1)\cdot N^2\right)$. 

When combining our iterative embedding process with TREXPIC, just like in the case of LE, we defined the required distance-like link weights with the simple formula 
\begin{equation}
    w_{ij}=1-\cos(\Delta\theta_{ij}).
    \label{eq:weightForTREXPIC}
\end{equation}
This way, we made TREXPIC placing nodes that lie at small angular distance $\Delta\theta_{ij}$ from each other hyperbolically closer in the subsequent iteration. Nevertheless, as it is clearly shown by the common approximating formula~\cite{hyperGeomBasics}
\begin{equation}
    x_{ij} \approx r_i+r_j+\frac{2}{\zeta}\cdot\ln\left(\frac{\Delta\theta_{ij}}{2}\right)
    \label{eq:hypDistApprox}
\end{equation}
derived from Eq.~(\ref{eq:hypLawOfCosines}), bringing nodes closer to each other hyperbolically goes hand in hand with decreasing their angular separation since the hyperbolic distance $x_{ij}$ is a decreasing function of the angular distance $\Delta\theta_{ij}$. Thus, during the iteration of TREXPIC, the communities that are usually arranged along the angular coordinates in the native representation of the hyperbolic space become more and more separated.

\subsection{Node embedding with Isomap}
\label{subsect:ISO}

The Isomap (ISO) method was originally proposed in Ref.~\cite{Isomap} for finding a lower-dimensional representation of data points given in a higher-dimensional space based on the eigendecomposition of the shortest path length (SPL) matrix of a neighborhood graph constructed from the data set given in the high-dimensional input space. Similarly to the case of Laplacian Eigenmaps (LE), the application of ISO for embedding real-world networks relies on the principle that the given graph can be considered as a neighborhood graph created from some high-dimensional data. Nevertheless, rather than trying to preserve local information and grasp specifically the direct connections between the nodes like LE, ISO aims at preserving the global network topology directly 
by searching for such a node arrangement in the $d$-dimensional Euclidean space in which the pairwise Euclidean distances between the nodes are as close to the topological distances --- given by the SPLs --- measured along the network as it is possible in a $d$-dimensional representation.

Throughout this study, instead of the original algorithm that works with eigendecomposition~\cite{Isomap}, we followed the implementation published for Ref.~\cite{coalescentEmbedding}, which is built on singular value decomposition (SVD), just like the above-described TREXPIC method. Thus, when embedding a network in the $d$-dimensional Euclidean space with ISO, we assigned to each network node a position vector of $d$ number of Cartesian coordinates through the following steps:
\begin{enumerate}
    \item Creation of the matrix $\mathbfit{D}$ of expected pairwise Euclidean distances between the network nodes in the embedding space: $D_{ij}\equiv D_{ji}=\mathrm{SPL}_{ij}$.
    \begin{enumerate}
        \item For unweighted graphs, $\mathrm{SPL}_{ij}$ is the possible smallest number of hops in which the $j$th network node can be reached from the $i^{\rm th}$ network node along the links.
        \item If link weights $w_{ij}\geq0$ are given, these are interpreted as the distance between the connected nodes $i,\,j$ in the hidden high-dimensional space, meaning that a larger link weight is read as a weaker connection between two nodes, and two nodes connected with a link of $0$ weight --- which must not be confused with an unconnected node pair --- can access each other the easiest. In this case, $\mathrm{SPL}_{ij}$ is the possible smallest sum of link weights along the paths that connect the $i^{\rm th}$ network node to the $j$th one.
    \end{enumerate}
    In both cases, $\mathrm{SPL}_{ii}=0$ for any node index $i$.
    \item Conversion of the matrix $\mathbfit{D}$ of expected Euclidean distances to the matrix $\mathbfit{I}$ of the corresponding expected Euclidean dot products. Choosing the origin as the position of the center of mass of the network nodes (which choice does not influence the distances between the nodes in the embedding), this is done by
    \begin{enumerate}
        \item[i.] constructing the matrix $\mathbfit{S}$ of squared expected distances from the elements $S_{ij}=D_{ij}^2$,
        \item[ii.] subtracting the corresponding averages from the rows and the columns of $\mathbfit{S}$ to create a doubly centered version $\mathbfit{S}_{\mathrm{dc}}$ where the mean of all rows and also all columns is $0$,
        \item[iii.] and finally, using the formula $\mathbfit{I}=-\mathbfit{S}_{\mathrm{dc}}/2$.
    \end{enumerate}
    \item Calculation of the $N\times d$-sized matrix $\mathbfit{Y}$ of Cartesian node coordinates that closely reproduce the expected dot products (and hereby also the expected Euclidean distances), i.e. that fulfills $\mathbfit{I}=\mathbfit{Y}\cdot\mathbfit{Y}^{\mathrm{T}}$ as much as it is viable using the given number of dimensions. As it was described in Ref.~\cite{HOPE}, when searching for $d$-dimensional position vectors, then an optimal solution for minimizing the L2-norm of $\mathbfit{I}-\mathbfit{Y}\cdot\mathbfit{Y}^T$ is to 
    \begin{enumerate}
        \item[i.] find the largest $d$ number of singular values ${\sigma_1\geq\sigma_2\geq...\geq\sigma_d}$ and the corresponding singular vectors $\underline{u}_1,\,\underline{u}_2,\,...,\,\underline{u}_d$ of the dot product matrix~$\mathbfit{I}$
        \item[ii.] and compute the $\ell^{\rm th}$ ($\ell=1,\,2,\,...,\,d$) Cartesian coordinate in the position vector of the $i^{\rm th}$ ($i=1,\,2,\,...,\,N$) network node, i.e. the element in the $i^{\rm th}$ row and $\ell^{\rm th}$ column of the coordinate matrix $\mathbfit{Y}$ as $\underline{y}_i(\ell)=\sqrt{\sigma_{\ell}}\cdot\underline{u}_{\ell}(i)$.
    \end{enumerate}
    Note that in the $N$-dimensional Euclidean space, the $N\times N$-sized coordinate matrix calculated according to the above steps can be written based on the singular value decomposition (SVD) of the dot product matrix $\mathbfit{I}=\mathbfit{U}\cdot\mathbfit{\Sigma}\cdot\mathbfit{V}^{\mathrm{T}}$ (where $\mathbfit{\Sigma}$ is the diagonal matrix of singular values, and the columns of the matrixes $\mathbfit{U}$ and $\mathbfit{V}$ are the left and the right singular vectors of $\mathbfit{I}$, respectively) as $\mathbfit{Y}=\mathbfit{U}\cdot\sqrt{\mathbfit{\Sigma}}$ or, since for symmetric $\mathbfit{I}$ matrixes obtained from undirected networks $\mathbfit{U}=\mathbfit{V}$, as $\mathbfit{Y}=\mathbfit{V}\cdot\sqrt{\mathbfit{\Sigma}}$, yielding $\mathbfit{Y}\cdot\mathbfit{Y}^{\mathrm{T}}=\mathbfit{U}\cdot\sqrt{\mathbfit{\Sigma}}\cdot\left(\mathbfit{V}\cdot\sqrt{\mathbfit{\Sigma}}\right)^{\mathrm{T}}=\mathbfit{U}\cdot\sqrt{\mathbfit{\Sigma}}\cdot\sqrt{\mathbfit{\Sigma}}\cdot\mathbfit{V}^{\mathrm{T}}\equiv\mathbfit{I}$, meaning that at $d=N$ the $N\times N$-sized dot product matrix $\mathbfit{I}$ is fully reproduced.
\end{enumerate}
The computational complexity of embedding $N$ network nodes in the $d$-dimensional Euclidean space with the above algorithm is dominated at ${d<\ln(N)}$ by the calculation of all the shortest path lengths in the network, yielding $\pazocal{O}\left(\ln(N)\cdot N^2\right)$, while the computational complexity of the truncated SVD is $\pazocal{O}\left(d\cdot N^2\right)$.

To enable capturing the topological relations of a network more precisely, we developed a new version of ISO that we call Isomap with exponentialized shortest path lengths. Here, following the idea of LE and TREXPIC (see Eqs.~(\ref{eq:LEdistProxConv}) and (\ref{eq:expSPLforTREXPIC}), respectively), we applied a relatively fast-changing, exponential function on the original topological measures, 
namely instead of using simply $D_{ij}\equiv D_{ji}=\mathrm{SPL}_{ij}$, we defined the expected Euclidean distances in step 1 of the above algorithm of ISO as
\begin{equation}
    D_{ij}=e^{-\,\frac{t}{\mathrm{SPL}_{ij}}}\in[0,1),
    \label{eq:expSPLforISO}
\end{equation}
just like we calculated the expected hyperbolic distances from the pairwise shortest path lengths between the nodes in Eq.~(\ref{eq:expSPLforTREXPIC}). For the parameter $t$, we used the default setting applied in TREXPIC, namely ${t=\sqrt{\ln(1.0/0.9999)\cdot\ln(1.0/0.1)}\cdot\mathrm{SPL}_{\mathrm{max}}}$, where $\mathrm{SPL}_{\mathrm{max}}$ denotes the largest shortest path length occurring in the examined network. 
When using our IERW process with the exponentialized ISO method, we re-set the $t$ parameter in each iteration according to the current $\mathrm{SPL}_{\mathrm{max}}$ yielded by the current link weights.

Note that by exponentializing in Eq.~(\ref{eq:expSPLforISO}) not directly the link weights but the measured shortest path lengths instead, the effect of the exponentialization becomes visible in the angular arrangement of the embedded nodes of not only weighted but also unweighted networks. In unweighted networks, where all the link weights are the same (e.g. $1$), the exponentialization of each link weight individually corresponds to simply multiplying all the SPLs in the network with the same constant factor, which results in a simple rescaling of all the pairwise expected inner products that leads to a multiplication of all the singular values of the matrix $\mathbfit{I}$ with a constant factor but does not change the direction of the position vectors. Therefore, to utilize the exponentialization in unweighted networks too, we opted in Eq.~(\ref{eq:expSPLforISO}) for exponentializing not the individual link weights but the shortest path lengths. 

The impact of the exponentialization of the SPLs in ISO is demonstrated by Sect.~\ref{sect:expVSnotExp}, showing that this modification greatly enhances the increase in the communities' angular separation ratio $\langle \Delta\theta \rangle_{\rm inter}/\langle \Delta\theta \rangle_{\rm intra}$ during the embedding iterations and our iterative embedding process with the exponentialized version of Isomap can be successful in contracting the communities even with the application of the really simple distance-like link weight formula 
\begin{equation}
    w_{ij}=1-\cos(\Delta\theta_{ij})
    \label{eq:weightForExpSIO}
\end{equation}
also used in the case of iterating LE and TREXPIC (see Eqs.~(\ref{eq:weightForLE}) and (\ref{eq:weightForTREXPIC}), respectively). 

\subsection{Node embedding with node2vec}
\label{subsect:node2vec}

The node2vec method~\cite{node2vec} generates $d$-dimensional Euclidean embeddings based on the presumption that the network topology can be thoroughly explored via random walks. In this method, the local environment around a node is described through multiple truncated random walks along the links that yield ordered node lists having a given, restricted length. These node lists are used for embedding the nodes by following the same idea that was utilized in the word2vec algorithm~\cite{word2vec} for creating spatial arrangements of words representing their similarities from the point of view of their typical contexts in given texts. node2vec replaces the "sentences" (or, more precisely, the context windows, i.e., the considered lists of consecutive words) with lists of consecutive nodes within random walks, trains an artificial neural network to learn the characteristic neighborhood of the network nodes and assigns spatial positions with small distances to those node pairs that frequently appear close to each other in the random paths. In our study, following Ref.~\cite{commDetInEmbs_Santo}, we always set the number of walks started from each node to $80$, the length of each random walk to $10$ and the length of the considered windows to $10$.

The random walkers follow different node-node transition probabilities in node2vec, controlled by the return parameter $p$ and the in-out parameter $q$, setting the non-normalized transition probabilities to
\begin{itemize}
    \item $1/p$ for stepping back to that node from which the walker just arrived to the current node,
    \item $1$ for stepping to a common neighbor of the current node and the previous one,
    \item and $1/q$ for the other cases, i.e. for moving away ("outward") from the preceding node.
\end{itemize}
Low values of $p$ and high values of $q$ bias the walks to be more local, proceeding less to farther nodes from a given starting node, resulting in embeddings that preserve well the explored neighborhood relations, while at high values of $p$ and low values of $q$ node2vec generates node arrangements focusing more on the global network topology. In our measurements, we always used the settings $p=1$ and $q=1$.

If link weights are given in the network to be embedded, node2vec modifies the transition probabilities accordingly: here a larger link weight is interpreted as a stronger connection, yielding a higher transition probability between the given two nodes. A simple choice for proximity-like link weights based on embeddings obtained from node2vec in our iterative embedding process is given by
\begin{equation}
    w_{ij}=\cos(\Delta\theta_{ij})+1,
    \label{eq:notExpWeightFornode2vec}
\end{equation} 
where $\Delta\theta_{ij}$ denotes the angular distance between the $i^{\rm th}$ and the $j$th network nodes and the addition of $1$ ensures that all the link weights are non-negative (namely, $w_{ij}\in[0,2]$). However, following the idea of Laplacian Eigenmaps and TREXPIC about including an exponentialization step in the embedding process, to emphasize the differences between the simple link weights given by Eq.~(\ref{eq:notExpWeightFornode2vec}), we applied a rather fast-changing function on these, and defined exponential link weights in the IERW process as 
\begin{equation}
    w_{ij}=e^{t\cdot\left(\cos(\Delta\theta_{ij})-1\right)},
    \label{eq:expWeightFornode2vec}
\end{equation} 
where we replaced the original $+1$ term with a $-1$ in order to exponentialize non-positive values instead of non-negative ones and avoid numerical (overflow) errors this way. Note that the exponential link weights given by Eq.~(\ref{eq:expWeightFornode2vec}) are still non-negative (fall in the range $[e^{-2t},1]$), and thus, can be interpreted as transition probabilities. We show the advantageous effect of using the exponential link weight formula in Eq.~(\ref{eq:expWeightFornode2vec}) instead of Eq.~(\ref{eq:notExpWeightFornode2vec}) in Sect.~\ref{sect:expVSnotExp}. Assuming that the heterogeneity of the degree distribution plays an important role from the viewpoint of the transition probabilities in random walks and that it can be well described by the ratio between the average node degree $\bar{\kappa}$ and the most frequent node degree, i.e. the mode $\hat{\kappa}$ of the degrees, we set the parameter $t$ to $t=10\cdot\bar{\kappa}/\hat{\kappa}$. If the mode of the degrees was not unique, we set $\hat{\kappa}$ to the smallest one of the most frequently occurring node degrees. Unlike in the case of LE (Eq.~(\ref{eq:LEdistProxConv})), TREXPIC (Eq.~(\ref{eq:expSPLforTREXPIC})) and the exponentialized version of ISO (Eq.~(\ref{eq:expSPLforISO})), where we re-set the $t$ parameter in each iteration according to the current link weights, in the case of node2vec (Eq.~(\ref{eq:expWeightFornode2vec})) we used the same value of $t$ at each iteration for a given network, namely the one that we chose based on the original, unweighted graph, considering only the number of connections and not node strengths. 

The time complexity of node2vec for embedding a network of $N$ nodes and $E$ edges into a $d$-dimensional space is $\pazocal{O}\left(E+N\cdot d\cdot\omega^2\right)$, where $\omega$ denotes the window length, which was set to $\omega=10$ in our measurements. It is important to note that due to the application of random walks, node2vec is a stochastic embedding method, yielding different node arrangements when re-run for the same network. Nevertheless, since we repeated each of our measurements for multiple network samples anyway, we ran IERW with node2vec only once for each network.

\newpage
\renewcommand{\thesection}{S2}
\section{The beneficial effect of exponentialization in Isomap and node2vec}
\label{sect:expVSnotExp}
\setcounter{subsection}{0}

As it is shown by Eqs.~(\ref{eq:LEdistProxConv}) and (\ref{eq:expSPLforTREXPIC}) in Sect.~\ref{sect:embAlgorithms}, both Laplacian Eigenmaps~\cite{LE} and TREXPIC~\cite{TREXPIC} perform an exponentializing step when embedding a network. Expecting that such relatively fast-changing measures of the topological proximity and distance can effectively emphasize the differences between the relations of different node pairs and thereby improve the embedding performance, we wanted to include some exponentialization in the proposed Iterative Embedding and ReWeighting (IERW) process in the case of Isomap~\cite{Isomap} and node2vec~\cite{node2vec} too. Therefore, as it is detailed in Sects.~\ref{subsect:ISO} and \ref{subsect:node2vec}, we created a modified version of Isomap by defining the expected Euclidean distances in Eq.~(\ref{eq:expSPLforISO}) as exponentialized shortest path lengths, and included the exponentialization in the iteration of node2vec by choosing for it in IERW an exponential link weight function given by Eq.~(\ref{eq:expWeightFornode2vec}).

This section demonstrates the beneficial effect of introducing
exponentialization in the IERW process in the case of Isomap and node2vec by comparing the angular separation ratios achieved with the exponentialized and the non-exponentialized versions on the networks examined in Figs.~\ref{fig:angSepRate_PP} and \ref{fig:angSepRate_LFR} of the main text. In the case of Isomap, the not exponential version corresponds to the original algorithm~\cite{Isomap}, where the expected Euclidean distances are defined as the simple shortest path lengths. In the case of IERW with node2vec, we defined the non-exponentialized version by replacing the exponential link weight function in the reweighting steps with the linear link weight function given by Eq.~(\ref{eq:notExpWeightFornode2vec}).

\begin{figure}[!h]
    \centering
    \includegraphics[width=0.9\textwidth]{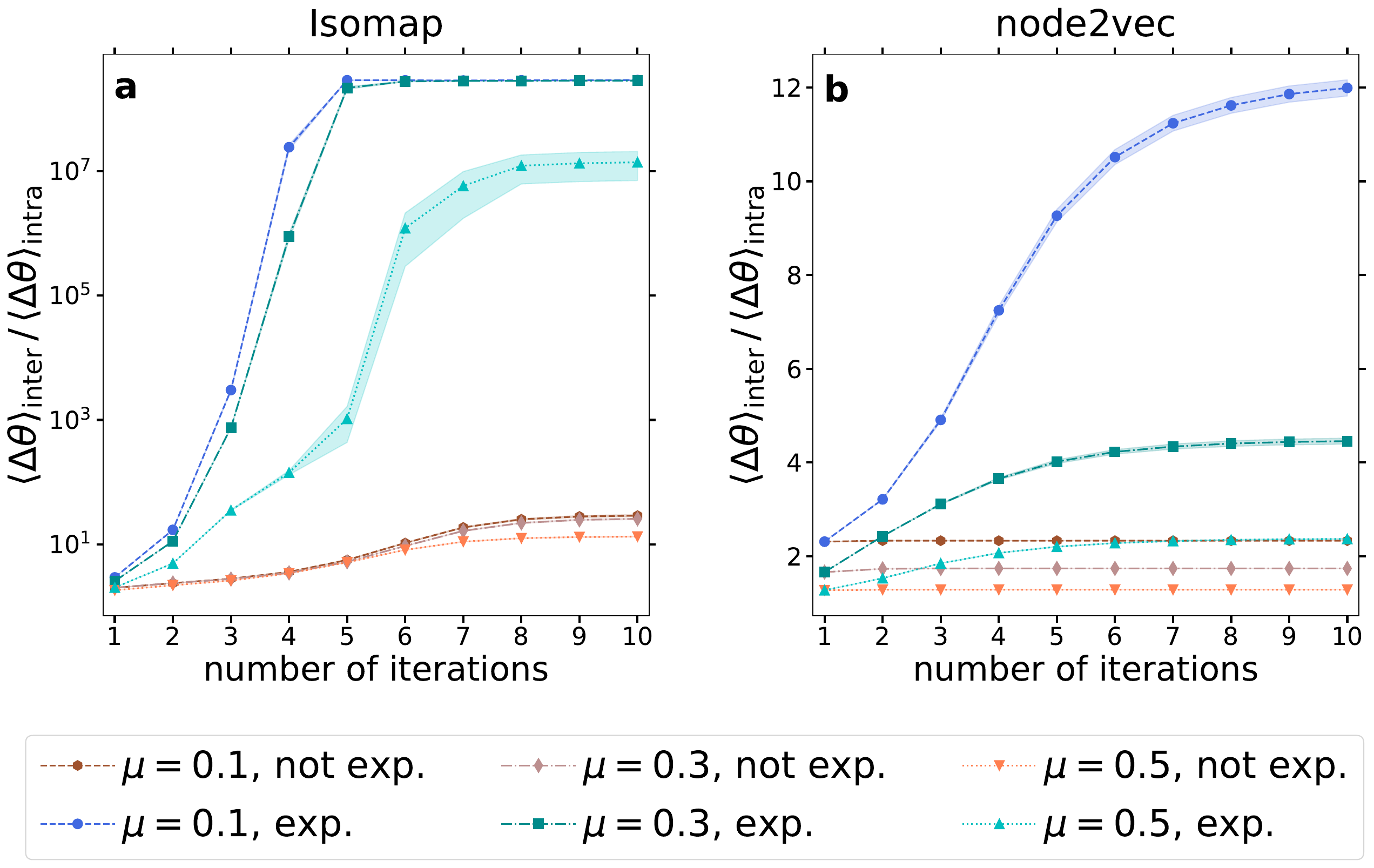}
    \caption{ {\bf The beneficial effect of exponentialization in the iteration of Isomap (panel \textbf{a}) and node2vec (panel \textbf{b}) on the increase in the angular separation of the communities in planted partition networks.} $\langle \Delta\theta \rangle_{\rm inter}$ denotes the average of the angular distances over all possible node pairs of different communities, while $\langle \Delta\theta \rangle_{\rm intra}$ stands for the average of the angular distances over all possible node pairs belonging to the same community. The curves of different line styles correspond to different values of the mixing parameter $\mu$, yielding community structures of different detectability levels. The blue and green curves that refer to the exponentialized versions are the same as in panels \textbf{c} and \textbf{d} of Fig.~\ref{fig:angSepRate_PP} in the main text. The network generation parameters are given in the Methods section of the main text. Each depicted data point corresponds to the result averaged over $100$ networks of the given parameter setting and the shaded areas show the standard error of the means.}
    \label{fig:angSepRate_expVSnotExp_PP1000}
\end{figure}

\begin{figure}[!h]
    \centering
    \includegraphics[width=0.9\textwidth]{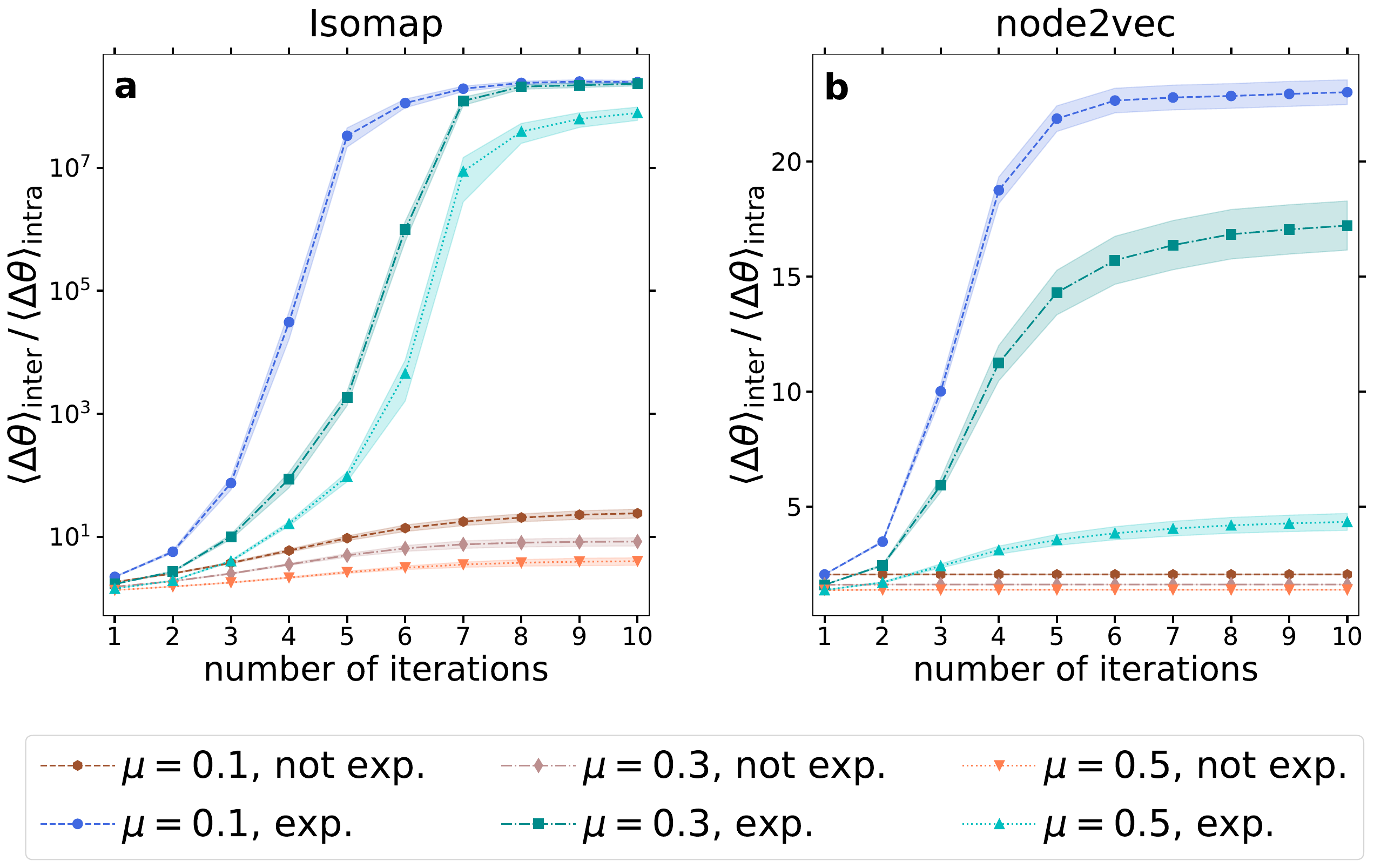}
    \caption{ {\bf The beneficial effect of exponentialization in the iteration of Isomap (panel \textbf{a}) and node2vec (panel \textbf{b}) on the increase in the angular separation of the communities in Lancichinetti--Fortunato--Radicchi networks.} $\langle \Delta\theta \rangle_{\rm inter}$ denotes the average of the angular distances over all possible node pairs of different communities, while $\langle \Delta\theta \rangle_{\rm intra}$ stands for the average of the angular distances over all possible node pairs belonging to the same community. The curves of different line styles correspond to different values of the mixing parameter $\mu$, yielding community structures of different detectability levels. The blue and green curves that refer to the exponentialized versions are the same as in panels \textbf{c} and \textbf{d} of Fig.~\ref{fig:angSepRate_LFR} in the main text. The network generation parameters are given in the Methods section of the main text. Each depicted data point corresponds to the result averaged over $100$ networks of the given parameter setting and the shaded areas show the standard error of the means.}
    \label{fig:angSepRate_expVSnotExp_LFR1000}
\end{figure}

Figure~\ref{fig:angSepRate_expVSnotExp_PP1000} supplements Fig.~\ref{fig:angSepRate_PP} of the main text, comparing the exponentialized and the non-exponentialized embedding iterations on planted partition networks~\cite{plantedPartitionModel}. Figure~\ref{fig:angSepRate_expVSnotExp_LFR1000} supplements Fig.~\ref{fig:angSepRate_LFR} of the main text, showing the same comparison for Lancichinetti--Fortunato--Radicchi networks~\cite{LFRbenchmark}. The advantage of the exponentialized versions is striking for both types of test networks. Note that in the case of Isomap, at the smaller mixing parameters the first embedding already performs slightly better when using the exponentialization, which is due to our choice of exponentializing not the individual link weights (that are initially all equal to $1$) but the shortest path lengths instead.

\newpage
\renewcommand{\thesection}{S3}
\section{Embedding parameters}
\label{sect:embParams}
\setcounter{subsection}{0}

This section deals with the settings of the parameters of the examined four embedding methods. First, Sect.~\ref{subsect:chooseEmbDim} describes in detail how the number of embedding dimensions was set. Then, Sect.~\ref{subsect:chooseTfactor} considers the other parameter that can be tuned in all the cases, namely the scaling factor in the exponentializing step. Finally, Sect.~\ref{subsect:measuringTheEffectOfEmbParams} demonstrates that the applied parameter choices are (although typically not the optimal but) relatively favorable, falling in a reasonable range of the parameter space.

\subsection{Choosing the number of embedding dimensions}
\label{subsect:chooseEmbDim}

In the case of node2vec, according to our measurements shown in Sect.~\ref{subsect:measuringTheEffectOfEmbParams}, the optimal number of embedding dimensions for revealing the community structure is not related strongly to the number of communities planted in a network, and the main requirement of node2vec is simply using a high enough number of dimensions. Therefore, to ensure a substantial reduction in the number of embedding dimensions $d$ compared to the number of network nodes but also utilize the benefits of a relatively high-dimensional space, we always used node2vec with the setting $d=64$, which is one of the standard choices in the literature.

Nevertheless, based on the measurements presented in Sect.~\ref{subsect:measuringTheEffectOfEmbParams}, the matrix decomposition methods given by LE, TREXPIC and ISO are more sensitive to the number of dimensions of the embedding space. In general, it can be stated that node embeddings can grasp more information and thereby describe the local peculiarities of a network topology more precisely if the number of dimensions of the embedding space is higher. However, when aiming at the exploration of the communities of the network nodes, i.e. when focusing on the network structure at a mesoscopic scale instead of the individual pairwise interactions, using an excessive number of dimensions can be problematic too. When performing a dimension reduction of a matrix that characterizes the network topology, one has to separate from each other the important components and those that are rather redundant from the viewpoint of the given task. We assume that the number of variables necessary for properly describing the community structure is independent of the specific type of the matrix that is reduced in the embedding. 

To find the proper number of variables from the viewpoint of community detection, let us consider the eigendecomposition $\mathbfit{L}_{\mathrm{norm}}\cdot\underline{g}_{\ell}=\lambda_{\ell}\cdot\underline{g}_{\ell}$ of the normalized Laplacian matrix
\begin{equation}
    \mathbfit{L}_{\mathrm{norm}}=\boldsymbol{\pazocal{D}}^{-1/2}\cdot\mathbfit{L}\cdot\boldsymbol{\pazocal{D}}^{-1/2}=\boldsymbol{\pazocal{D}}^{-1/2}\cdot(\boldsymbol{\pazocal{D}}-\mathbfit{A})\cdot\boldsymbol{\pazocal{D}}^{-1/2}
    \label{eq:normLaplacian}
\end{equation} 
of an undirected graph with an adjacency matrix $\mathbfit{A}$ and a diagonal matrix $\boldsymbol{\pazocal{D}}$, with $\pazocal{D}_{ii}=\sum_j A_{ij}$. It is well known that here the smallest eigenvalue $\lambda_{\mathrm{min}}$ is always $0$, and the most meaningful components are given by the eigenvectors belonging to the smallest non-zero eigenvalues. The question is how many eigenvectors should be retained, separating them from the less relevant eigenvectors that describe only local peculiarities. Considering simply the largest gap between consecutive elements in the increasing order of the (non-zero) eigenvalues as the point of transition from the truly meaningful "small" (but non-zero) eigenvalues to the less important "large" ones, the number of non-zero eigenvalues below this transition point can be a good proxy for the proper number of variables to be retained for revealing the community structure of a network. 
The idea that this selection between the eigenvalues may be suitable for grasping the main properties of the network topology in relation to the communities is supported by the fact --- described e.g. in Ref.~\cite{eigengapsOfGraphLaplacian} --- that while the multiplicity of the $0$ eigenvalue is equal to the number of connected components in the graph (which is $1$ in our measurements), the number of eigenvalues being either $0$ or close to $0$ is the same as the number of groups the network nodes can be partitioned into via sparse cuts, indicating that the number of small eigenvalues of the matrix $\mathbfit{L}_{\mathrm{norm}}$ and the number of communities in a network are strongly connected to each other. 

\begin{figure}[!h]
    \centering
    \includegraphics[width=0.8\textwidth]{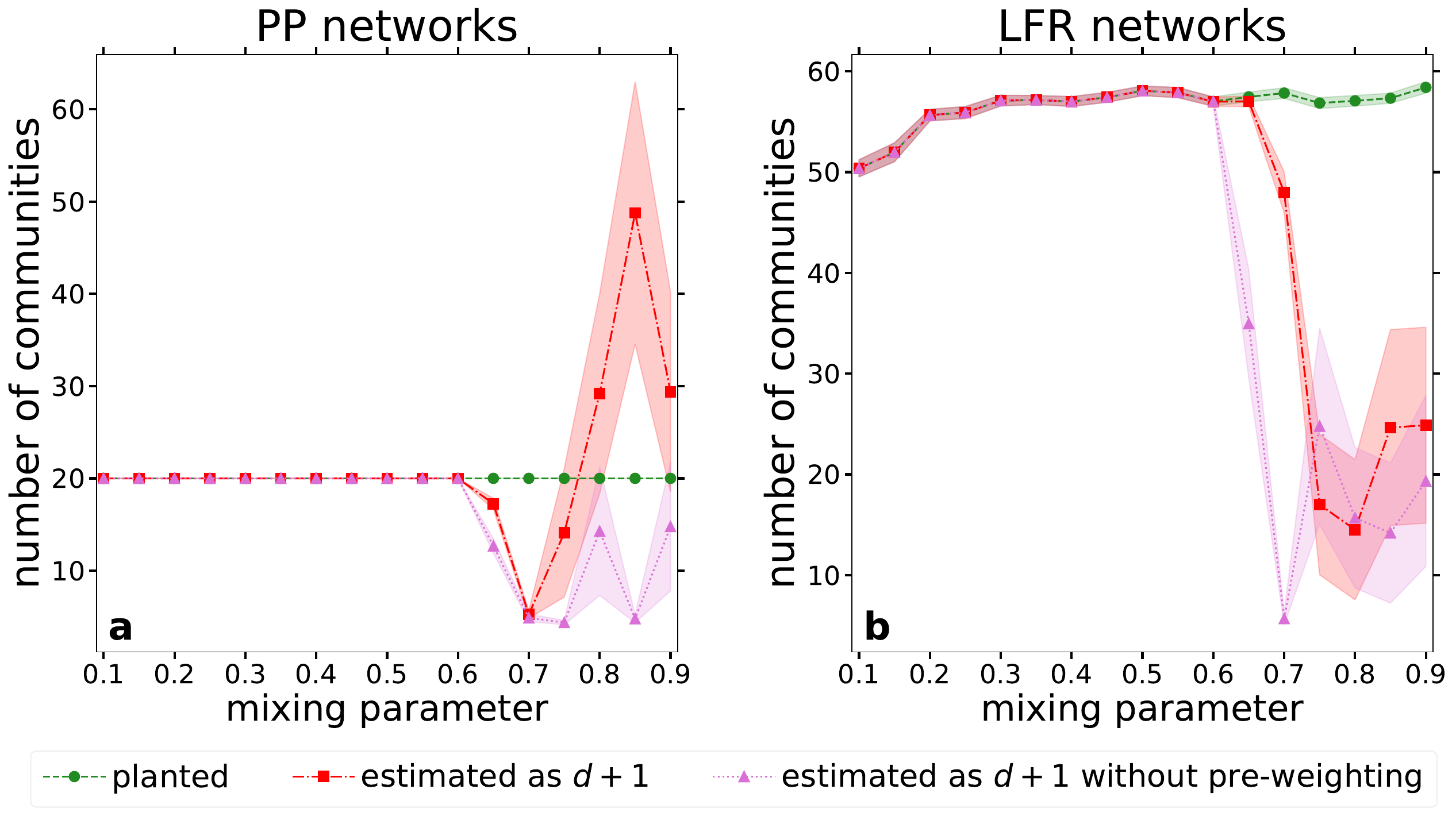}
    \caption{ {\bf The relation between the number of embedding dimensions chosen by the above algorithm and the number of planted communities in synthetic networks.} Panel \textbf{a} examines the planted partition graphs that were also studied e.g. in Fig.~\ref{fig:angSepRate_PP} of the main text, and panel \textbf{b} deals with the Lancichinetti--Fortunato--Radicchi networks that were also investigated in Fig.~\ref{fig:angSepRate_LFR} of the main text. To confirm the beneficial effect of the pre-weighting step, the $d$ values chosen based on the eigengaps of the graph Laplacian of the original, not weighted graphs are depicted too. Each data point corresponds to the average over $100$ networks of the given parameter settings and the error bars show the standard error of the mean.
    }
    \label{fig:chooseDim_connectionWithNumOfComms}
\end{figure}

Therefore, to set the proper number of embedding dimensions for LE, TREXPIC and ISO, we simply considered all the non-zero eigenvalues of a normalized graph Laplacian falling below the largest gap in the increasing order of the non-zero eigenvalues to be important. More specifically, we chose the number of embedding dimensions $d$ of LE, TREXPIC and ISO through the following steps:
\begin{enumerate}
    \item Pre-weight the graph to facilitate the selection of $d$. 
    \begin{enumerate}
        \item[i.] Assign a distance-like weight to each edge $i-j$ in the network following a repulsion-attraction rule (RA1) proposed in Ref.~\cite{coalescentEmbedding} for describing neighborhood topological information and the trade-off of the attraction between nodes having a high number of common neighbors and the repulsion between nodes having a lot of connections in distinct neighborhoods: 
        \begin{equation}
            w_{ij}=\frac{\kappa_i+\kappa_j+\kappa_i\cdot\kappa_j}{1+\mathrm{CN}_{ij}},
            \label{eq:RA1}
        \end{equation}
        where $\kappa_i$ and $\kappa_j$ stand for the total number of links connected to node $i$ and node $j$, respectively, and $\mathrm{CN}_{ij}$ denotes the number of common neighbors of nodes $i$ and $j$. Here a large link weight can emerge due to the presence of a relatively high number of neighbors not in common, indicating a more vulnerable connection or a larger topological distance.
        \item[ii.] Convert the obtained distance-like link weights to proximity-like ones using the formula applied in the LE algorithm~\cite{LE} (see Eq.~(\ref{eq:LEdistProxConv}))
        \begin{equation}
            w_{ij}'(w_{ij})=e^{-\frac{w_{ij}^2}{t}},
            \label{eq:distProxConvInDimSelection}
        \end{equation} 
        setting the scaling factor $t$ to the square of the mean of the weights, as it was suggested in the LE implementation created for Ref.~\cite{coalescentEmbedding}. 
    \end{enumerate}
    When working with a weighted graph, instead of the above-described artificial link weights, the actual weights can also be utilized. If all weights are non-negative and higher values mean stronger or more relevant connections, these can be used without modification. If, however, higher link weights correspond to larger topological distances, these can be converted to proximity-like link weights e.g. with the exponential formula given by Eq.~(\ref{eq:distProxConvInDimSelection}). Note that we do not use the pre-weights to do the embeddings, but only to set the number of dimensions of the embedding space.
    \item Create the normalized Laplacian matrix $\mathbfit{L}_{\mathrm{norm}}$ of the pre-weighted graph according to Eq.~(\ref{eq:normLaplacian}).
    \item Compute the smallest $\lfloor N/2\rfloor+1$ number of eigenvalues of the normalized graph Laplacian $\mathbfit{L}_{\mathrm{norm}}$, where the choice $\lfloor N/2\rfloor+1$ sets the allowed largest number of embedding dimensions to $d_{\mathrm{max}}=\lfloor N/2\rfloor$ and thereby ensures a significant dimension reduction compared to the total number of network nodes $N$.
    \item Sort the eigenvalues in increasing order, omitting the $0$ eigenvalue to ensure that only the gaps between the non-zero eigenvalues will be taken into account. 
    \item Find the largest gap between consecutive eigenvalues in the obtained ordered list of $\lfloor N/2\rfloor$ elements and perform a Modified Thompson Tau test with a significance level of $\alpha=10^{-6}$ to decide whether the largest gap size can be considered to be an outlier among all the examined differences of successive eigenvalues.    
    \begin{enumerate}
        \item If the largest gap size can be considered as an outlier, and thus, as a significant eigengap, set the number of embedding dimensions $d$ to the number of elements of the examined list below the largest eigengap, yielding $d\in[1,\lfloor N/2\rfloor-1]$. If we get $d=1$, we use $d=2$.
        \item Otherwise, consider the largest gap to be irrelevant and --- assuming that there is no significant community structure in the given network --- use $d=d_{\mathrm{max}}$.
    \end{enumerate}
\end{enumerate}
As it is demonstrated by Fig.~\ref{fig:chooseDim_connectionWithNumOfComms}, for not too large values of the mixing parameter $\mu$, the number of dimensions $d$ chosen by the above algorithm is a good indicator of the number of communities $C$ planted in the investigated synthetic networks, namely usually the chosen number of embedding dimensions was $d=C-1$. This fits the intuition since e.g. to describe all the pairwise relations between three communities, a two-dimensional pattern (i.e., a triangle) is needed in general. 

\subsection{Choosing the scaling factor in the exponentializing step}
\label{subsect:chooseTfactor}

As it is explained in Sect.~\ref{sect:embAlgorithms}, all four examined cases of IERW include an exponentializing step, either during the node embedding (when using LE, TREXPIC or the exponentialized version of ISO) or in the link weighting phase (in the case of IERW with node2vec). All the applied exponential formulas contain a tunable parameter given by the scaling factor~$t$.

First, the algorithm of Laplacian Eigenmaps starts with an exponential conversion of the inputted distance-like link weights $w_{ij}$ to proximity-like ones as $w_{ij}'=e^{-w_{ij}^2/t}$ (see Eq.~(\ref{eq:LEdistProxConv}) in Sect.~\ref{subsect:LE}). Here, larger values of the parameter $t>0$ blur the differences between the inputted distances more. At the extreme, $t=\infty$ treats all links equally, regardless of the corresponding distances. 
If not mentioned otherwise, we followed the implementation created for Ref.~\cite{coalescentEmbedding}, and calculated the scaling factor $t$ in LE as the square of the mean of the inputted distance-like weights. 

Second, TREXPIC builds on an exponential distance, using $e^{-t/\mathrm{SPL}_{ij}}\in[0,1)$ as the expected hyperbolic distance between nodes $i$ and $j$ (see Eq.~(\ref{eq:expSPLforTREXPIC}) in Sect.~\ref{subsect:TREXPIC}). Here, the multiplying factor $t>0$ controls the speed of the increase in the expected hyperbolic distance with the increase in the shortest path length (SPL). For small enough values of the parameter~$t$, $e^{-t/\mathrm{SPL}_{ij}}\approx 1-t/\mathrm{SPL}_{ij}$ and all the non-zero SPLs are converted to expected hyperbolic distances close to $1$, while the increase in $t$ shifts the non-unit expected distances (corresponding to finite SPLs) towards $0$. With the intention of avoiding any extreme settings, we followed Ref.~\cite{TREXPIC} and calculated the default value of $t$ from the occurring largest shortest path length $\mathrm{SPL}_{\mathrm{max}}$ as $t=\sqrt{\ln(1.0/0.9999)\cdot\ln(1.0/0.1)}\cdot\mathrm{SPL}_{\mathrm{max}}$, corresponding to the geometric mean of two extreme settings given by $t_{\mathrm{small}}=\ln(1.0/0.9999)\cdot\mathrm{SPL}_{\mathrm{max}}$ (yielding a largest expected hyperbolic distance of $0.9999$) and $t_{\mathrm{large}}=\ln(1.0/0.1)\cdot\mathrm{SPL}_{\mathrm{max}}$ (yielding a largest expected hyperbolic distance of $0.1$). 

In the case of Isomap, we introduced an exponetialization mimicking TREXPIC, setting the expected Euclidean distance between nodes $i$ and $j$ to the exponential distance $e^{-t/\mathrm{SPL}_{ij}}\in[0,1)$ (see Eq.~(\ref{eq:expSPLforISO}) in Sect.~\ref{subsect:ISO}). Since using exactly the same exponential formula, we transferred the default $t$ value from TREXPIC to the exponentialized version of Isomap too.

Finally, when iterating node2vec, we utilized exponentialization not during the embedding but in the link weighting step and calculated the proximity-like link weights after each embedding from the angular distances $\Delta\theta_{ij}\in[0,\pi]$ as $w_{ij}=e^{t\cdot\left(\cos(\Delta\theta_{ij})-1\right)}$ (see Eq.~(\ref{eq:expWeightFornode2vec}) in Sect.~\ref{subsect:node2vec}). Here, the multiplying factor $t>0$ controls the speed of the decrease in the random walk transition probability with the increase in the angular distance: at larger values of $t$, the transition probabilities decay faster as a function of $\Delta\theta_{ij}$. Assuming that the heterogeneity of the number of connections per node (i.e., the number of possible directions in which a random walk can be continued) plays an important role from the viewpoint of the transition probabilities in random walks, if not mentioned otherwise, we set the parameter $t$ to $t=10\cdot\bar{\kappa}/\hat{\kappa}$, where $\bar{\kappa}$ is the average node degree and $\hat{\kappa}$ is the most frequent node degree. If the mode of the degrees was not unique, we set $\hat{\kappa}$ to the smallest one of the most frequently occurring node degrees.

\subsection{Validating the choice of the embedding parameters}
\label{subsect:measuringTheEffectOfEmbParams}

\begin{figure}[!h]
    \centering
    \includegraphics[width=1.0\textwidth]{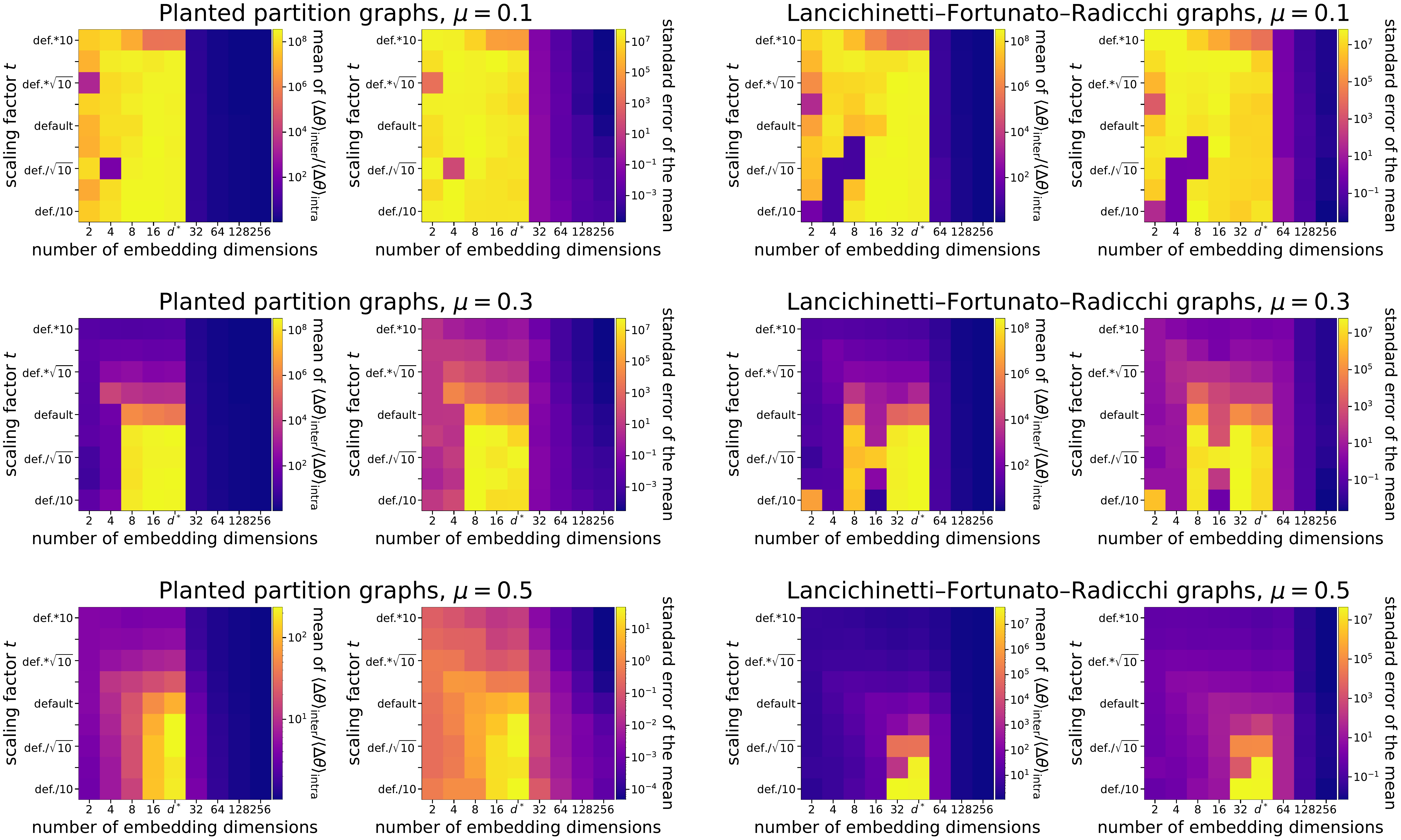}
    \caption{ {\bf The effect of the embedding parameters on the ratio between the average inter-community angular distance $\langle \Delta\theta \rangle_{\rm inter}$ and the average intra-community angular distance $\langle \Delta\theta \rangle_{\rm intra}$ at the final step when iterating Laplacian Eigenmaps.} Each pair of subplots depicts the results for $10$ realizations of a given type of synthetic networks: the left half of the figure refers to planted partition graphs (also examined in Fig.~\ref{fig:angSepRate_PP} of the main text), while the panels on the right deal with Lancichinetti--Fortunato--Radicchi networks (also examined in Fig.~\ref{fig:angSepRate_LFR} of the main text). The IERW procedure was performed only once for each network. The default scaling factor was set according to Ref.~\cite{coalescentEmbedding}, and $d^*$ is the number of embedding dimensions chosen by our algorithm described in Sect.~\ref{subsect:chooseEmbDim}.
    }
    \label{fig:paramDependence_LE}
\end{figure}

\begin{figure}[!h]
    \centering
    \includegraphics[width=1.0\textwidth]{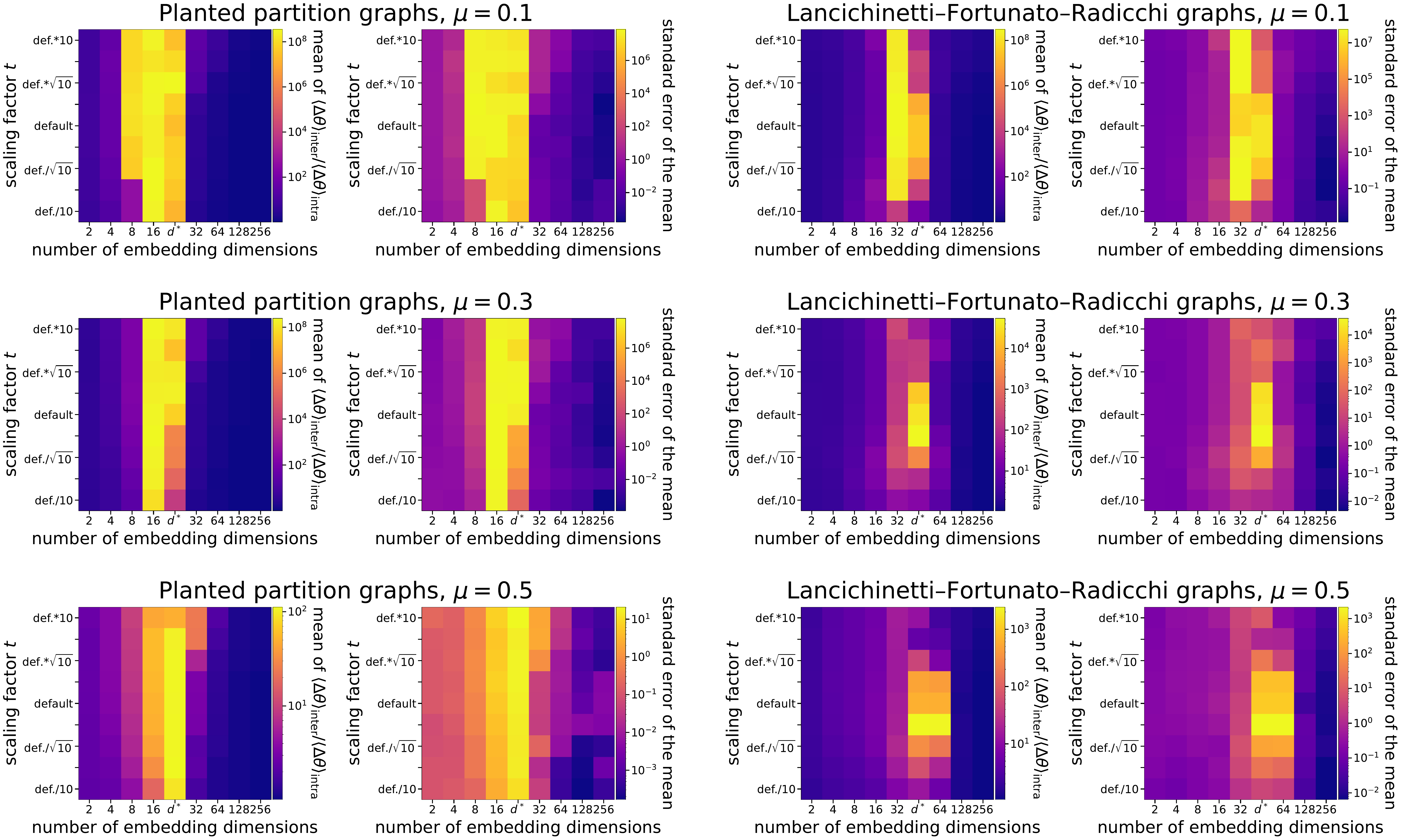}
    \caption{ {\bf The effect of the embedding parameters on the ratio between the average inter-community angular distance $\langle \Delta\theta \rangle_{\rm inter}$ and the average intra-community angular distance $\langle \Delta\theta \rangle_{\rm intra}$ at the final step when iterating TRansformation of EXponential shortest Path lengths to hyperbolIC measures.} Each pair of subplots depicts the results for $10$ realizations of a given type of synthetic networks: the left half of the figure refers to planted partition graphs (also examined in Fig.~\ref{fig:angSepRate_PP} of the main text), while the panels on the right deal with Lancichinetti--Fortunato--Radicchi networks (also examined in Fig.~\ref{fig:angSepRate_LFR} of the main text). The IERW procedure was performed only once for each network. The default scaling factor was set according to Ref.~\cite{TREXPIC}, and $d^*$ is the number of embedding dimensions chosen by our algorithm described in Sect.~\ref{subsect:chooseEmbDim}.
    }
    \label{fig:paramDependence_TREXPIC}
\end{figure}

\begin{figure}[!h]
    \centering
    \includegraphics[width=1.0\textwidth]{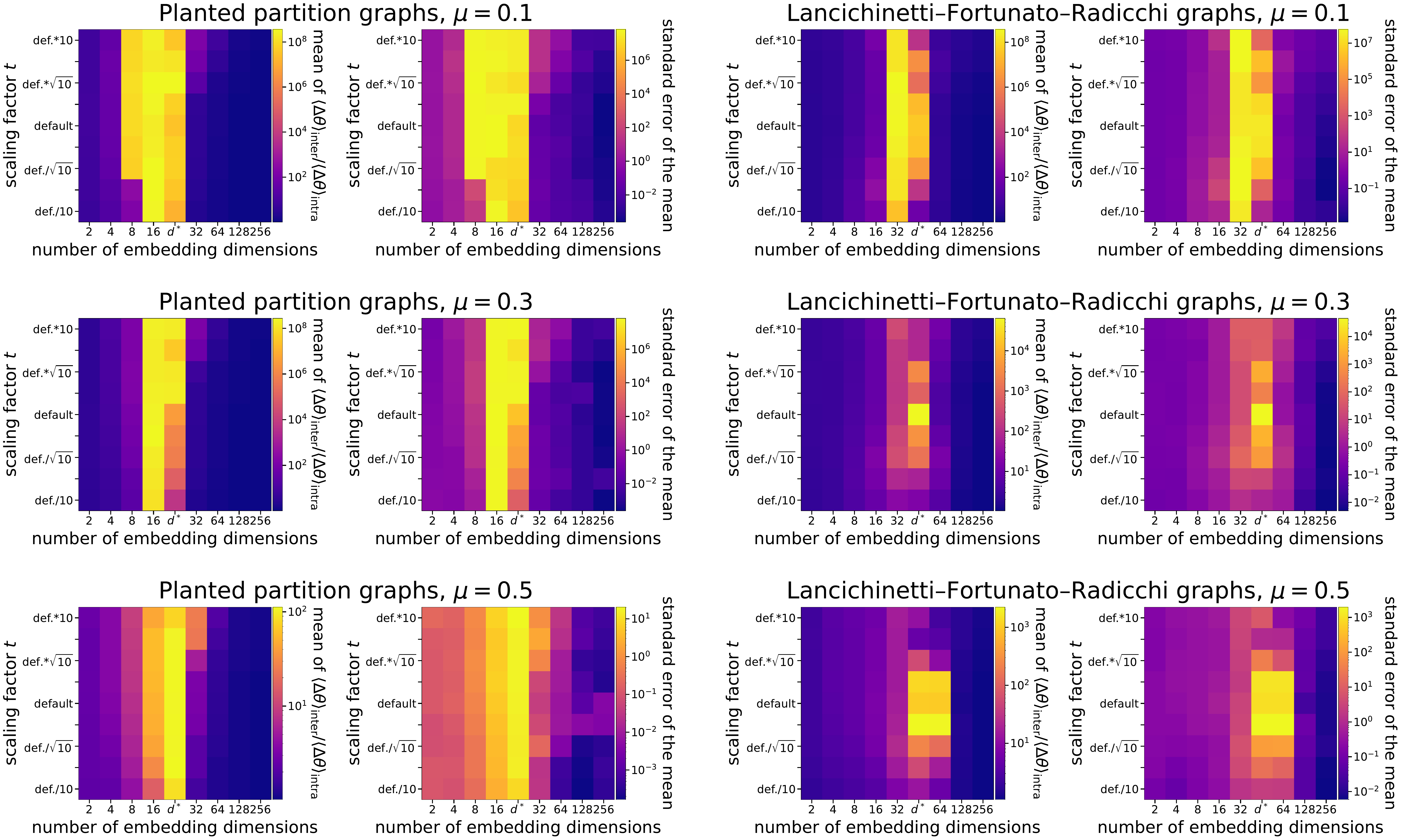}
    \caption{ {\bf The effect of the embedding parameters on the ratio between the average inter-community angular distance $\langle \Delta\theta \rangle_{\rm inter}$ and the average intra-community angular distance $\langle \Delta\theta \rangle_{\rm intra}$ at the final step when iterating Isomap with exponentialized shortest path lengths.} Each pair of subplots depicts the results for $10$ realizations of a given type of synthetic networks: the left half of the figure refers to planted partition graphs (also examined in Fig.~\ref{fig:angSepRate_PP} of the main text), while the panels on the right deal with Lancichinetti--Fortunato--Radicchi networks (also examined in Fig.~\ref{fig:angSepRate_LFR} of the main text). The IERW procedure was performed only once for each network. The default scaling factor was set according to our formula defined in Sect.~\ref{subsect:chooseTfactor}, and $d^*$ is the number of embedding dimensions chosen by our algorithm described in Sect.~\ref{subsect:chooseEmbDim}.
    }
    \label{fig:paramDependence_expISO}
\end{figure}

\begin{figure}[!h]
    \centering
    \includegraphics[width=1.0\textwidth]{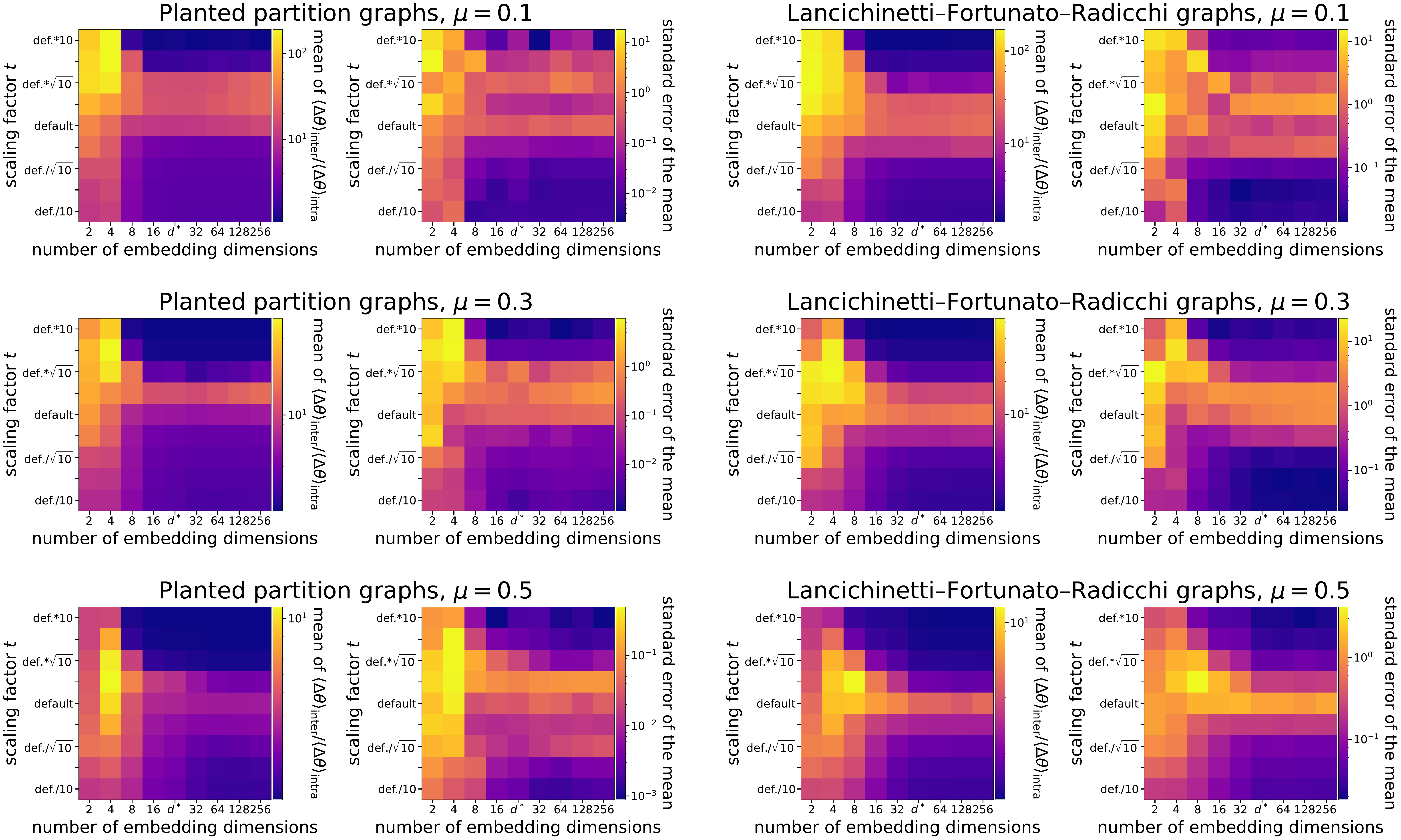}
    \caption{ {\bf The effect of the embedding parameters on the ratio between the average inter-community angular distance $\langle \Delta\theta \rangle_{\rm inter}$ and the average intra-community angular distance $\langle \Delta\theta \rangle_{\rm intra}$ at the final step when iterating node2vec with exponentialized link weights.} Each pair of subplots depicts the results for $10$ realizations of a given type of synthetic networks: the left half of the figure refers to planted partition graphs (also examined in Fig.~\ref{fig:angSepRate_PP} of the main text), while the panels on the right deal with Lancichinetti--Fortunato--Radicchi networks (also examined in Fig.~\ref{fig:angSepRate_LFR} of the main text). The IERW procedure was performed only once for each network. The default scaling factor was set according to our formula defined in Sect.~\ref{subsect:chooseTfactor}, and $d^*$ is the number of embedding dimensions chosen by our algorithm described in Sect.~\ref{subsect:chooseEmbDim}.
    }
    \label{fig:paramDependence_expNode2vec}
\end{figure}

Figures \ref{fig:paramDependence_LE}-\ref{fig:paramDependence_expNode2vec} demonstrate how the change in the number of embedding dimensions $d$ and the scaling parameter $t$ of the exponentializing step affects the angular separation achieved between communities in iterated embeddings. In all the figures, $d^*$ denotes the number of embedding dimensions chosen according to the algorithm described in Sect.~\ref{subsect:chooseEmbDim} (approximating $C-1$, where $C$ is the number of communities planted in the given test network), and the "default" scaling factor corresponds to the setting described in Sect.~\ref{subsect:chooseTfactor}. We used our usual stopping criterion at all the settings, meaning that the iteration was terminated when the relative change in the average link weight between subsequent iterations dropped below $0.001$. Note that the number of iterations was limited to $20$ in order to reduce the computational time. At the most extreme parameter settings, sometimes the matrix factorization has failed due to numerical errors --- in these cases, we used the last successful embedding. After reaching the final embedding, we measured the ratio between the average inter-community angular distance $\langle \Delta\theta \rangle_{\rm inter}$ (i.e., the average of the angular distances over all the node pairs of different communities) and the average intra-community angular distance $\langle \Delta\theta \rangle_{\rm intra}$ (i.e., the average of the angular distances over all the node pairs belonging to the same community). 

Considering the number of embedding dimensions, when comparing Figs.~\ref{fig:paramDependence_LE}--\ref{fig:paramDependence_expISO} to Fig.~\ref{fig:paramDependence_expNode2vec}, it is conspicuous that the performance of LE, TREXPIC and the exponentialized version of ISO shows multiple orders of difference as a function of $d$, while (in the vicinity of the default scaling factor) node2vec shows relatively weak $d$-dependence along the whole examined range. According to Figs.~\ref{fig:paramDependence_LE}--\ref{fig:paramDependence_expISO}, our choice $d^*$ always fell in the range of high performance (even when this range was very tight), justifying the application of our $d$-selecting algorithm detailed in Sect.~\ref{subsect:chooseEmbDim}. Regarding the scaling factor $t$ in the exponentializing steps, our default settings seem to be adequate in general. 

\newpage
\mbox{}
\newpage
\mbox{}
\newpage
\mbox{}
\newpage
\renewcommand{\thesection}{S4}
\section{Improvement of the weight thresholding performance with the iteration of the embedding}
\label{sect:weightThreshold_firstVSfinalIter}
\setcounter{subsection}{0}

\begin{figure}[!h]
    \centering
    \includegraphics[width=1.0\textwidth]{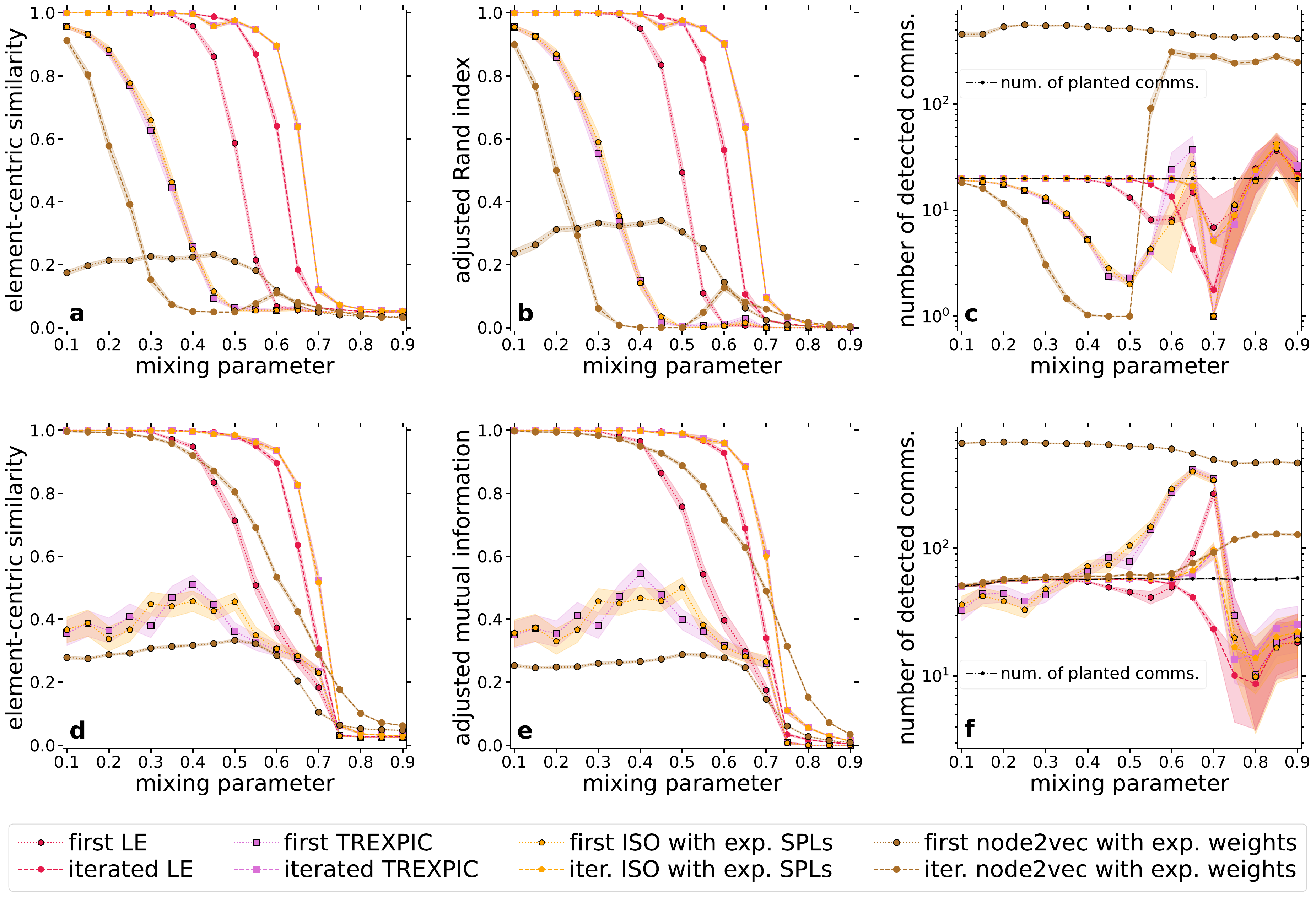}
    \caption{ {\bf Community detection performance of weight thresholding with one single embedding and with the iterated embedding in the synthetic networks 
    also examined in Fig.~\ref{fig:commDetWithWeightThreshold} of the main text.} The upper row (panels \textbf{a}, \textbf{b} and \textbf{c}) deals with the networks generated by the planted partition model, and the lower row (panels \textbf{d}, \textbf{e} and \textbf{f}) refers to the graphs yielded by the Lancichinetti--Fortunato--Radicchi benchmark. The dotted lines present the results obtained from a single embedding with Laplacian Eigenmaps (LE, red hexagons), TRansformation of EXponential shortest Path lengths to hyperbolIC measures (TREXPIC, purple squares), Isomap (ISO) with exponentialized shortest path lengths (orange pentagons) and node2vec with exponentialized link weights (brown circles), whereas the dashed lines show the results achieved by iterating the above embeddings. Each data point corresponds to the average over $100$ networks of the given parameter settings and the error bars show the standard error of the mean. We performed the community detection with all the methods only once for each network.}
    \label{fig:iterationHelpsInWeightThresholding_N1000}
\end{figure}

In the main text, Figs.~\ref{fig:angSepRate_PP} and \ref{fig:angSepRate_LFR} demonstrated that the angular separation between the communities planted in PP and LFR graphs can be significantly increased through the iteration of an embedding. Besides, Fig.~\ref{fig:commDetWithWeightThreshold} of the main text showed that by changing the link weights, iterated embeddings in many cases separate the inter-community links from the intra-community edges so well that they can make even a simple weight thresholding able to reveal the planted community structures with a quality comparable to that of well-known network community detection methods. Supplementing these results, Fig.~\ref{fig:iterationHelpsInWeightThresholding_N1000} confirms that the iteration of the embeddings is indeed necessary for achieving convincing community detection performances with the weight thresholding procedure described in the Methods section of the main text, as this rather crude approach in general performs very poorly when applied after only a single embedding.

\newpage
\renewcommand{\thesection}{S5}
\section{Improving traditional community detection methods with matrix factorization embedding methods}
\label{sect:tradCommDet_spectral_supp}
\setcounter{subsection}{0}

While Fig.~\ref{fig:tradCommDetHelpedWithExpNode2vec} in the main text demonstrated how IERW can be used with node2vec to improve the performance of traditional community detection methods, this section presents similar applications of iterating matrix factorization embedding techniques. Section~\ref{subsect:spectralNetworkCommDet} exemplifies through Louvain~\cite{Louvain,Louvain_code}, asynchronous label propagation~\cite{alabprop,alabprop_code} and Infomap~\cite{Infomap,Infomap_code} that well-known network community detection methods can effectively utilize the link weights obtained from IERW with LE, TREXPIC and ISO. Besides, Sect.~\ref{subsect:spectralHDBSCAN} illustrates through the example of Hierarchical Density-Based Spatial Clustering of Applications with Noise (HDBSCAN)~\cite{HDBSCAN1,HDBSCAN2,HDBSCAN_code} that standard spatial clustering methods can also benefit from iterating the LE, the TREXPIC and the ISO embedding methods.

\subsection{Enhancing network community detection using IERW with LE, TREXPIC and ISO}
\label{subsect:spectralNetworkCommDet}


As pointed out in the main text, while IERW with LE, TREXPIC and ISO embeddings yields distance-like link weights (where higher values indicate weaker connections or more distant relations), the Louvain~\cite{Louvain,Louvain_code}, the asynchronous label propagation~\cite{alabprop,alabprop_code} and the Infomap~\cite{Infomap,Infomap_code} community detection methods expect proximity-like link weights (where higher values indicate more intensive, stronger or closer relations). Therefore, before the application of Louvain, asynchronous label propagation or Infomap, one has to perform a conversion of the link weights provided by IERW with LE, TREXPIC and ISO. As mentioned in Eq.~(\ref{eq:distProxConvForTradMethods_main}) of the main text, this can be done e.g. by following a similar practice to the one suggested in Ref.~\cite{coalescentEmbedding} and using the conversion formula 
\begin{equation}
    \tilde{w}_{ij}=\frac{1}{w_{0}+w_{ij}},
    \label{eq:distProxConvForTradMethods}
\end{equation}
where $w_{0}>0$ is a tunable parameter. As $w_0$ decreases, the links with small distance-like weights $w_{ij}$ will be inputted to the traditional network community detection methods as stronger and stronger connections, while larger values of $w_0$ yield less difference between the resulting $\tilde{w}_{ij}$ proximity-like link weights of a network. 

\begin{figure}[!h]
    \centering
    \includegraphics[width=1.0\textwidth]{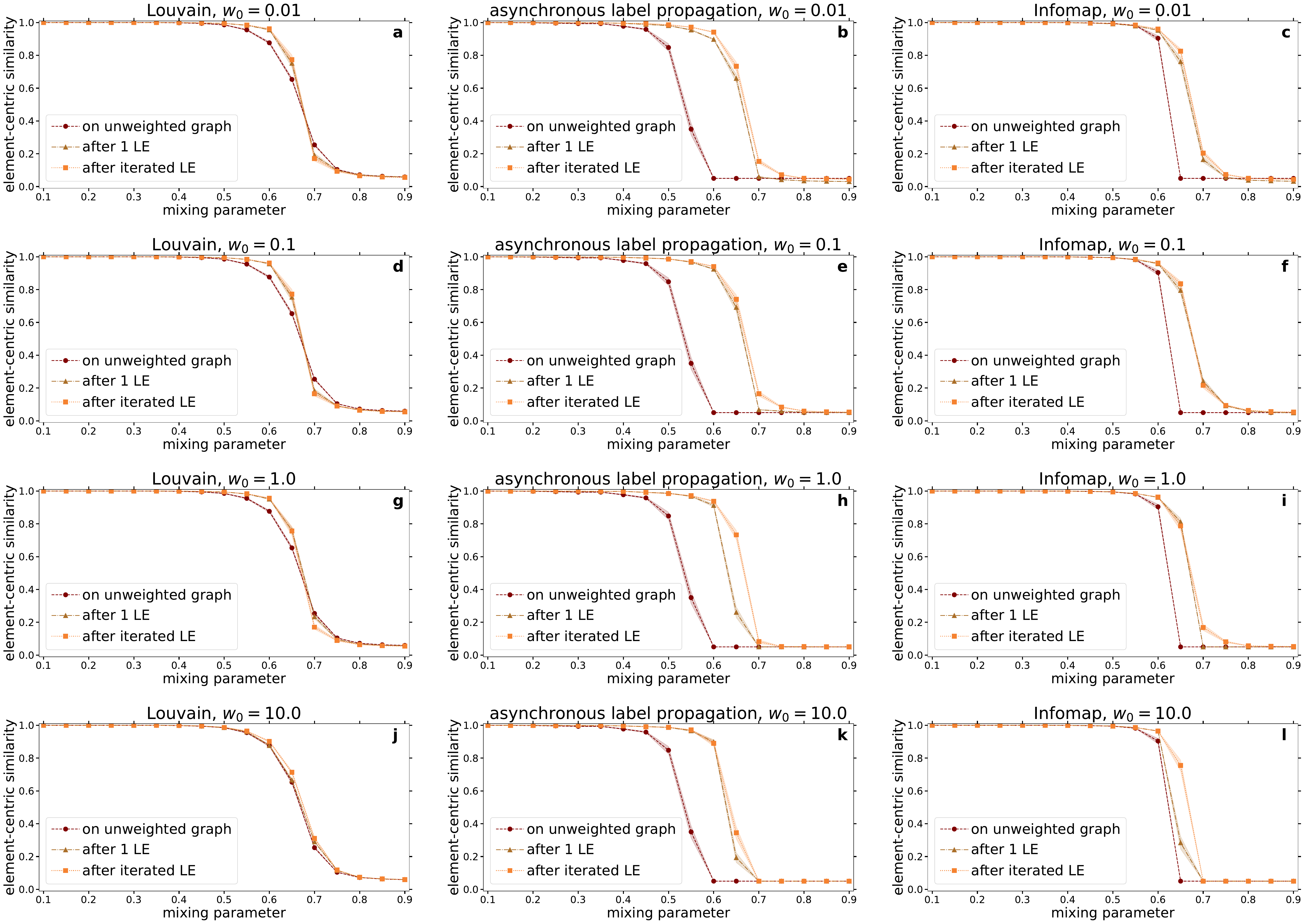}
    \caption{ {\bf Performance of usual network community detection methods without embedding, with one Laplacian Eigenmaps embedding and with iterated Laplacian Eigenmaps embedding in planted partition networks, using the embeddings with different $w_0$ values.} Each row of panels corresponds to a given setting of the $w_0$ parameter of Eq.~(\ref{eq:distProxConvForTradMethods}), and the different columns of panels refer to different community detection methods: Louvain, asynchronous label propagation and Infomap. We performed the community detection with all the examined methods only once for each network. Each displayed data point corresponds to a result averaged over $100$ networks, and the error bars depict the standard error of the mean.}
    \label{fig:tradCommDetHelpedWithLE_diffConsts_PP}
\end{figure}

\begin{figure}[!h]
    \centering
    \includegraphics[width=1.0\textwidth]{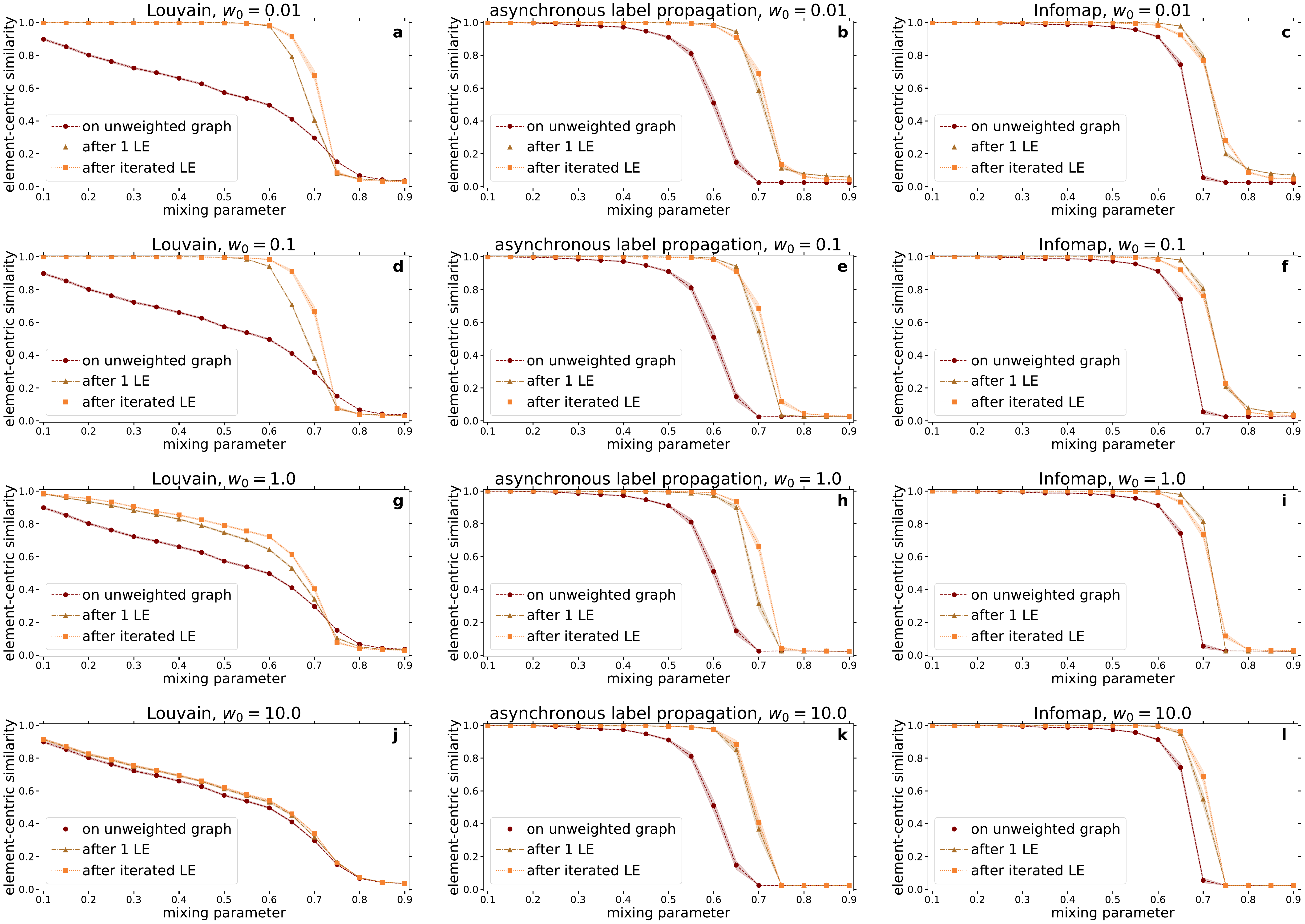}
    \caption{ {\bf Performance of usual network community detection methods without embedding, with one Laplacian Eigenmaps embedding and with iterated Laplacian Eigenmaps embedding in Lancichinetti--Fortunato--Radicchi networks, using the embeddings with different $w_0$ values.} Each row of panels corresponds to a given setting of the $w_0$ parameter of Eq.~(\ref{eq:distProxConvForTradMethods}), and the different columns of panels refer to different community detection methods: Louvain, asynchronous label propagation and Infomap. We performed the community detection with all the examined methods only once for each network. Each displayed data point corresponds to a result averaged over $100$ networks, and the error bars depict the standard error of the mean.}
    \label{fig:tradCommDetHelpedWithLE_diffConsts_LFR}
\end{figure}

Figures~\ref{fig:tradCommDetHelpedWithLE_diffConsts_PP} and \ref{fig:tradCommDetHelpedWithLE_diffConsts_LFR} demonstrate in the case of the LE embedding algorithm how the performance of the different network community detection methods depend on the $w_0$ parameter of the distance-proximity conversion formula given by Eq.~(\ref{eq:distProxConvForTradMethods}). According to the similarity scores obtained at the four different settings of $w_0$ on the same planted partition~\cite{plantedPartitionModel} (PP) and Lancichinetti--Fortunato--Radicchi~\cite{LFRbenchmark} (LFR) networks, Louvain --- which, due to the resolution limit of the modularity~\cite{resolution_limit_PNAS}, has a tendency of failing in detecting small communities~--- needs relatively small values of $w_0$ that put rather strong emphasis on the distance-like weights (i.e., cosine distances) that are close to $0$. For the PP networks at $w_0=10.0$ (Fig.~\ref{fig:tradCommDetHelpedWithLE_diffConsts_PP}j) and for the LFR networks at $w_0=1.0$ and $w_0=10.0$ (Fig.~\ref{fig:tradCommDetHelpedWithLE_diffConsts_LFR}g,j), the embedding-based link weights become unable to lead Louvain to the correct solution. Thus, for the Louvain method a relatively small $w_0$ seems to be the optimal choice, e.g. $w_0=0.1$.

On the other hand, it seems that small values of $w_0$ may highlight links of nearly $0$ distance-like weights so much that they make Infomap --- which has a much stronger tendency to find small communities than Louvain --- returning smaller groups of nodes within the planted communities instead of the whole planted communities. In such cases (Fig.~\ref{fig:tradCommDetHelpedWithLE_diffConsts_LFR}c,f,i), the first embedding provides more help than the iterated embedding for Infomap. This may be attributed to the emergence of some even denser patches within the relatively dense point clouds created by the first embedding from the planted communities, which is a natural consequence of the fact that due to the inhomogeneities of the connection structure, the spatial contraction of the individual communities is inhomogeneous during the embedding iteration. Therefore, the best solution for Infomap seems to be setting $w_0$ to a relatively large value of $10$, avoiding thereby the identification of the smaller subclusters of the planted groups as separate communities and making Infomap focusing on the larger point clouds that correspond to the planted communities. As an intermediate case between Louvain and Infomap, when using asynchronous label propagation, the iterative embedding has the largest advantage compared to a single (i.e., the first) embedding at $w_0=1.0$ (Figs.~\ref{fig:tradCommDetHelpedWithLE_diffConsts_PP}h and \ref{fig:tradCommDetHelpedWithLE_diffConsts_LFR}h). 

\begin{figure}[!h]
    \centering
    \includegraphics[width=1.0\textwidth]{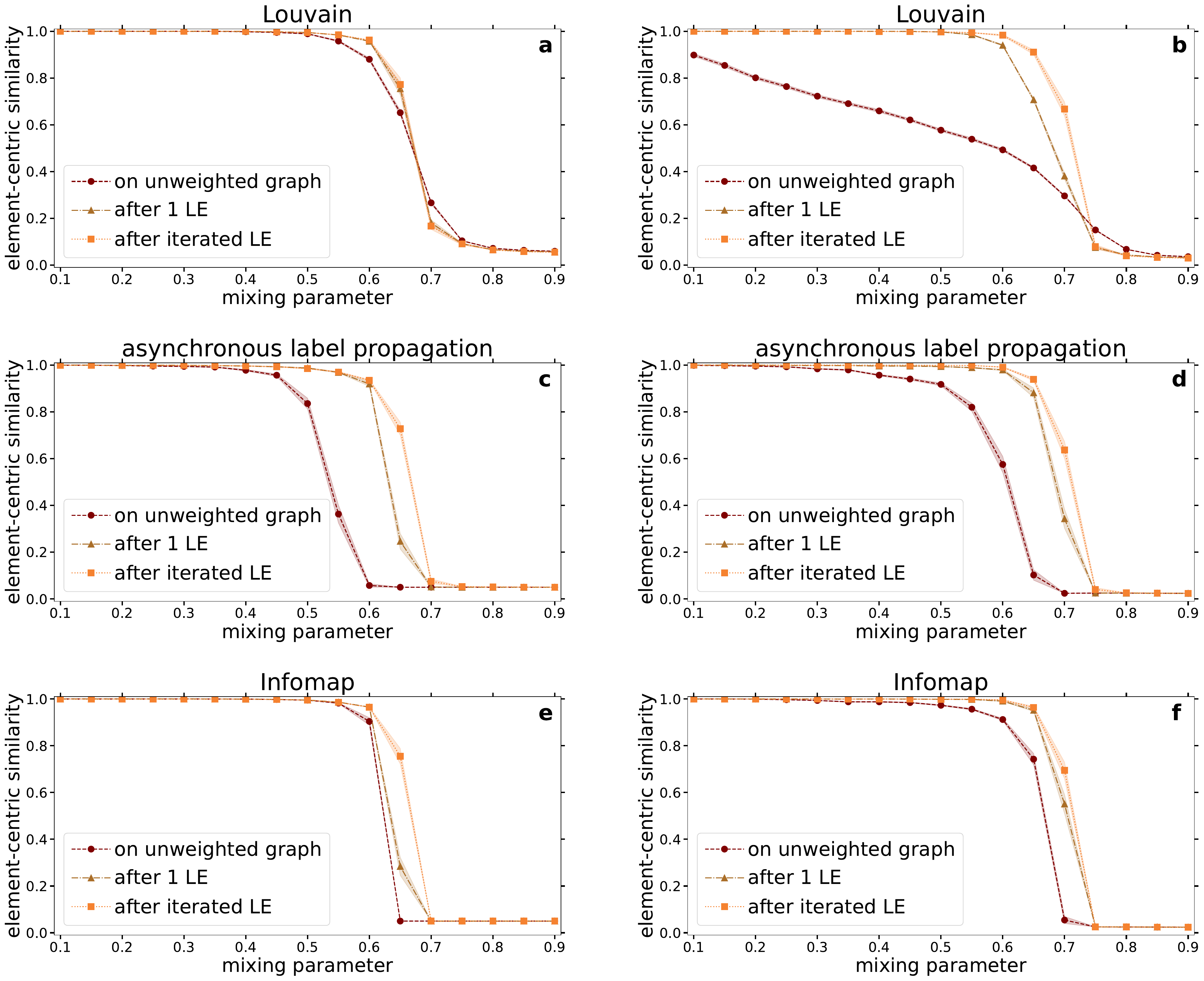}
    \caption{ {\bf Performance of usual network community detection methods on the weighted networks derived by IERW using Laplacian Eigenmaps and the "optimal" $w_0$ values.} Each row of panels corresponds to a different community detection method, and the left column refers to networks generated by the planted partition model, while the right one to networks generated by the Lancichinetti--Fortunato--Radicchi benchmark. We used here Eq.~(\ref{eq:distProxConvForTradMethods}) with the "optimal" settings of $w_0$ determined based on Figs.~\ref{fig:tradCommDetHelpedWithLE_diffConsts_PP} and \ref{fig:tradCommDetHelpedWithLE_diffConsts_LFR}, namely with $w_0=0.1$ in the case of Louvain (panels a and b), $w_0=1.0$ in the case of asynchronous label propagation (panels c and d) and $w_0=10.0$ in the case of Infomap (panels e and f). We performed the community detection with all the methods only once for each network. Each displayed data point corresponds to a result averaged over $100$ networks, and the error bars depict the standard error of the mean.}
    \label{fig:tradCommDetHelpedWithLE_optimalConsts}
\end{figure}

\begin{figure}[!h]
    \centering
    \includegraphics[width=1.0\textwidth]{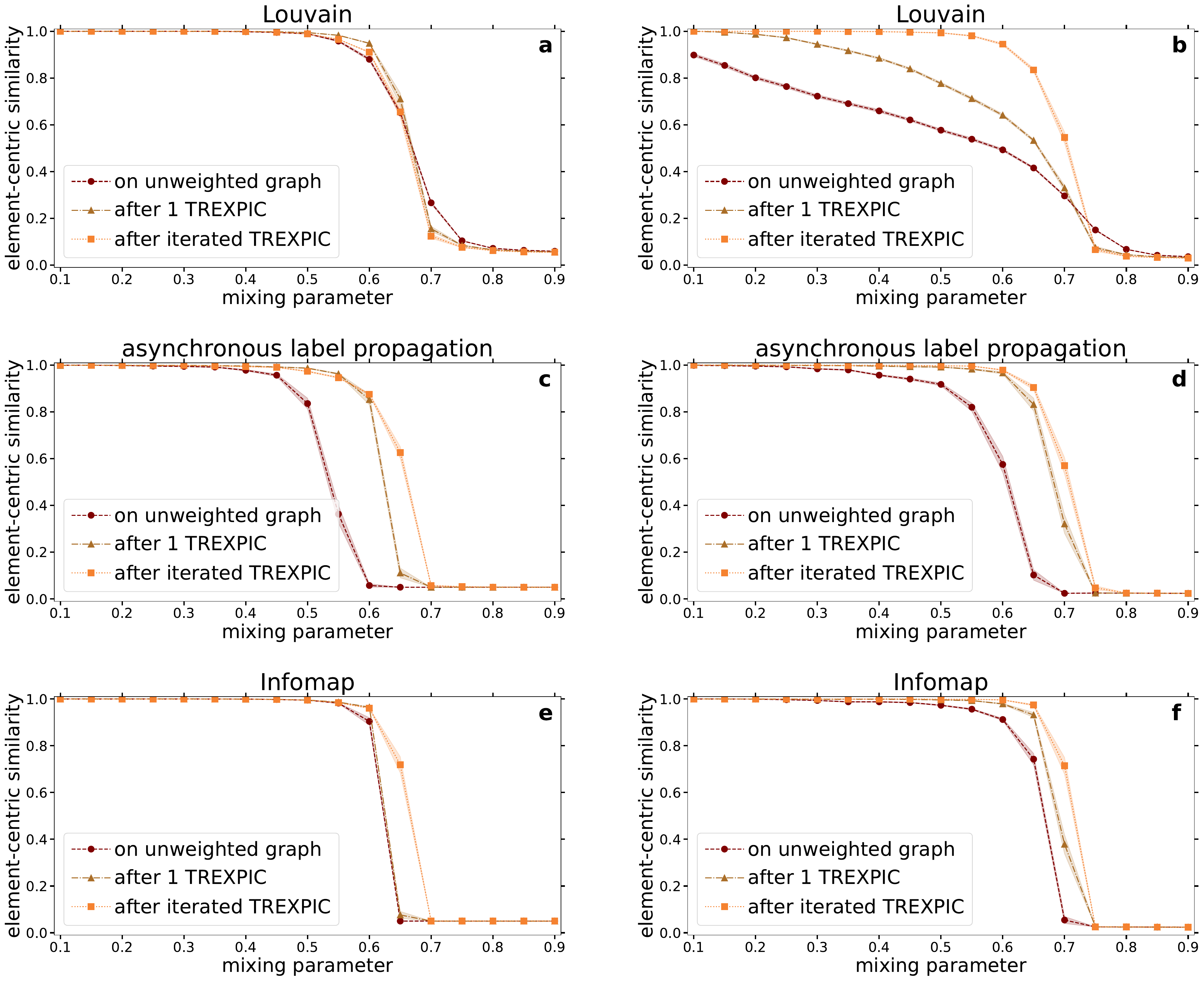}
    \caption{ {\bf Performance of usual network community detection methods on the weighted networks derived by IERW using TREXPIC and the "optimal" $w_0$ values.} The different rows of panels correspond to different community detection methods (named in the panel titles), and the left column of panels (i.e. panels \textbf{a}, \textbf{c} and \textbf{e}) refers to the networks generated by the planted partition model, while the right column of panels (i.e. panels \textbf{b}, \textbf{d} and \textbf{f}) deals with the networks obtained from the Lancichinetti--Fortunato--Radicchi benchmark. We used Eq.~(\ref{eq:distProxConvForTradMethods}) with the setting $w_0=0.1$ in the case of Louvain (panels a and b), $w_0=1.0$ in the case of asynchronous label propagation (panels c and d) and $w_0=10.0$ in the case of Infomap (panels e and f). We performed the community detection with all the examined methods only once for each network. Each displayed data point corresponds to a result averaged over $100$ networks, and the error bars depict the standard error of the mean.
    }
    \label{fig:tradCommDetHelpedWithTREXPIC_optimalConsts}
\end{figure}

\begin{figure}[!h]
    \centering
    \includegraphics[width=1.0\textwidth]{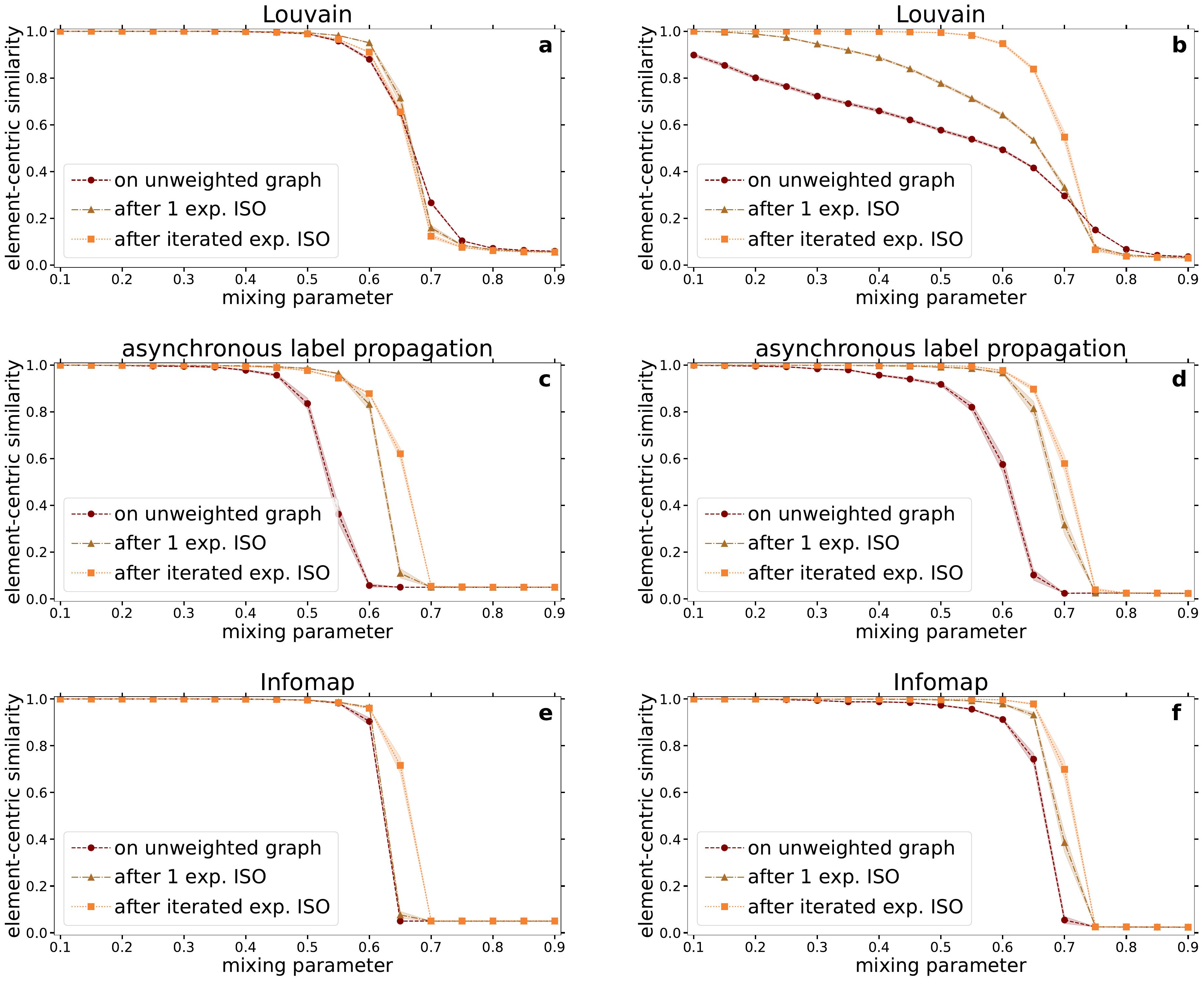} 
    \caption{ {\bf Performance of usual network community detection methods on the weighted networks derived by IERW using Isomap with exponentialized shortest path lengths and the "optimal" $w_0$ values.} The different rows of panels correspond to different community detection methods (named in the panel titles), and the left column of panels (i.e. panels \textbf{a}, \textbf{c} and \textbf{e}) refers to the networks generated by the planted partition model, while the right column of panels (i.e. panels \textbf{b}, \textbf{d} and \textbf{f}) deals with the networks obtained from the Lancichinetti--Fortunato--Radicchi benchmark. We used Eq.~(\ref{eq:distProxConvForTradMethods}) with the setting $w_0=0.1$ in the case of Louvain (panels a and b), $w_0=1.0$ in the case of asynchronous label propagation (panels c and d) and $w_0=10.0$ in the case of Infomap (panels e and f). We performed the community detection with all the examined methods only once for each network. Each displayed data point corresponds to a result averaged over $100$ networks, and the error bars depict the standard error of the mean.
    }
    \label{fig:tradCommDetHelpedWithExpISO_optimalConsts}
\end{figure}

After making our choices regarding the $w_0$ parameter of Eq.~(\ref{eq:distProxConvForTradMethods}) based on the examples shown in Figs.~\ref{fig:tradCommDetHelpedWithLE_diffConsts_PP} and \ref{fig:tradCommDetHelpedWithLE_diffConsts_LFR}, we tested the considered network community detection methods on the weighted graphs obtained from IERW with LE, TREXPIC and the exponentialized version of ISO, using $w_0=0.1$ in the case of Louvain, $w_0=1.0$ in the case of asynchronous label propagation and $w_0=10.0$ in the case of Infomap. Similarly to Fig.~\ref{fig:tradCommDetHelpedWithExpNode2vec}a--f in the main text that showed that IERW with node2vec can be used for facilitating these traditional network community detection methods, Figs.~\ref{fig:tradCommDetHelpedWithLE_optimalConsts}--\ref{fig:tradCommDetHelpedWithExpISO_optimalConsts} demonstrate that the link weights obtained from LE, TREXPIC or ISO in our IERW framework are capable of improving the performance of Louvain, asynchronous label propagation and Infomap. 

\subsection{Enhancing HDBSCAN using IERW with LE, TREXPIC and ISO}
\label{subsect:spectralHDBSCAN}

\begin{figure}[!h]
    \centering
    \includegraphics[width=1.0\textwidth]{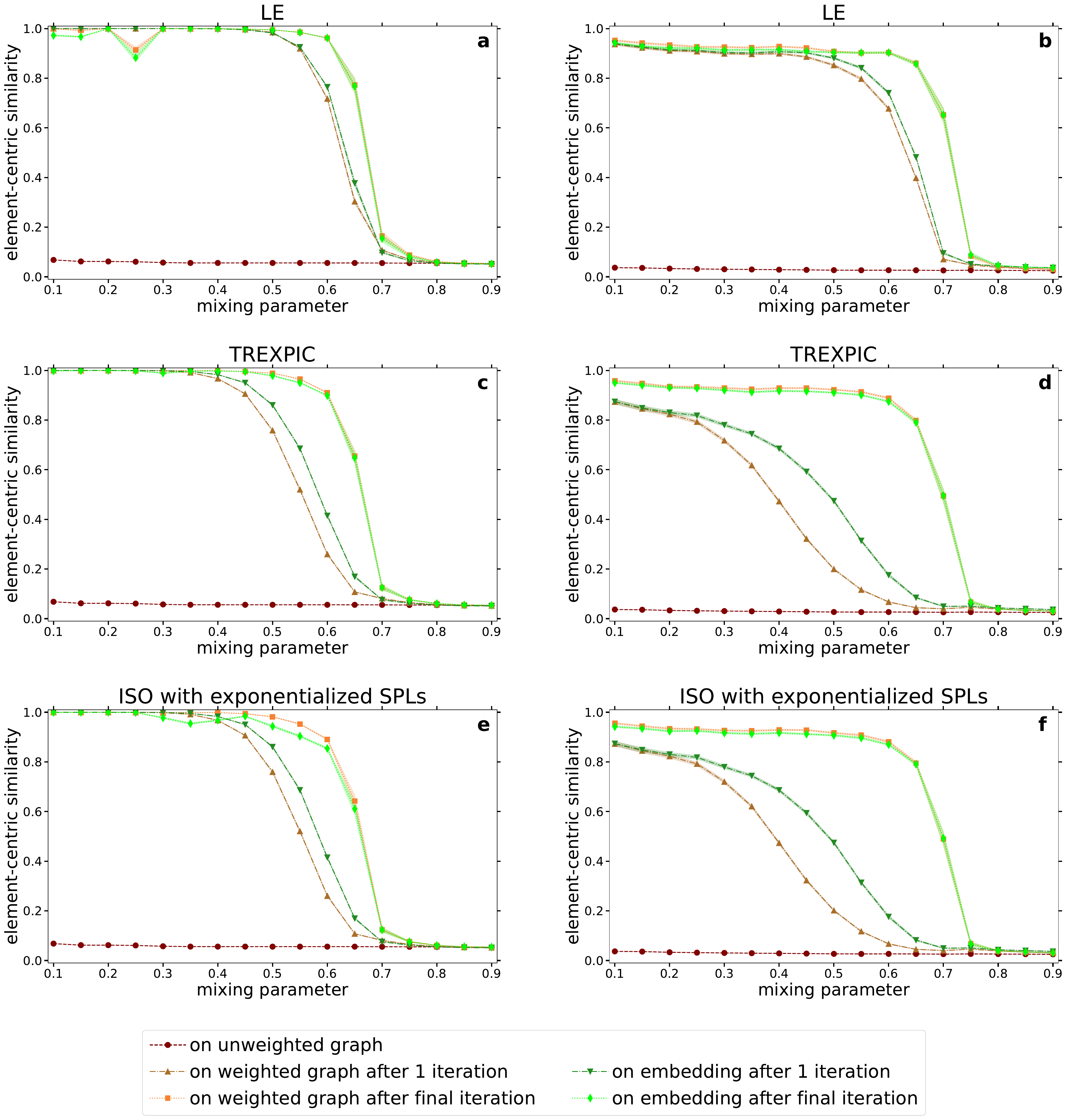}
    \caption{ {\bf Performance of HDBSCAN on the weighted networks and embeddings derived by IERW using embeddings based on matrix factorization.} The different rows of panels correspond to different embedding algorithms (named in the panel titles), and the left column of panels (i.e. panels \textbf{a}, \textbf{c} and \textbf{e}) refers to networks generated by the planted partition model, while the right column of panels (i.e. panels \textbf{b}, \textbf{d} and \textbf{f}) deals with the networks obtained from the Lancichinetti--Fortunato--Radicchi benchmark. In each panel, in the case of the two green curves, we used HDBSCAN on the matrix of cosine distances measured between the embedded nodes, while in the other three cases, we inputted into HDBSCAN the shortest path lengths measured along the network links. We performed all variants of the clustering only once for each network. Each displayed data point corresponds to a result averaged over $100$ networks, and the error bars depict the standard error of the mean.}
    \label{fig:HDBSCANhelpedWithSpectralEmbeddings}
\end{figure}

In the main text, Fig.~\ref{fig:tradCommDetHelpedWithExpNode2vec}g--h demonstrated that the iteration of node2vec embeddings (using exponential link weights) provides a significant boost to the spatial clustering method Hierarchical Density-Based Spatial Clustering of Applications with Noise (HDBSCAN)~\cite{HDBSCAN1,HDBSCAN2,HDBSCAN_code}. Here, Fig.~\ref{fig:HDBSCANhelpedWithSpectralEmbeddings} investigates the performance of HDBSCAN in relation to the examined matrix factorization embedding methods. Since HDBSCAN accepts any type of distance matrix as an input, we tested it using not only the usual pairwise geometric distances (measured directly in the embedding space, disregarding the connectedness of the nodes) but also distances measured along the network links. Namely, when running HDBSCAN on an embedding, we used the matrix of pairwise cosine distances $1-\cos(\Delta\theta)$ between the embedded nodes, and when running HDBSCAN on a graph, we inputted the matrix of shortest path lengths (SPLs) of the (possibly weighted) graph. According to our IERW procedure, the link weights were defined for all three embedding methods by the cosine distances between the connected nodes in the embedding (see Eqs.~(\ref{eq:weightForLE}), (\ref{eq:weightForTREXPIC}) and (\ref{eq:weightForExpSIO})). As it is shown by Fig.~\ref{fig:HDBSCANhelpedWithSpectralEmbeddings}, HDBSCAN was not able to find the planted communities based on the SPLs measured on the original, unweighted graphs (where each hop has the same contribution to the SPL and moving from one community to another can take only a single hop) even when the mixing between the communities was small. However, with the application of embedding-based link weights, the inter- and intra-community distances measured along the network links become much more distinguishable, raising the performance of the graph-based HDBSCAN to a level comparable to that of the embedding-based HDBSCAN. While the embedding-based HDBSCAN still seems to perform better after a single embedding, when the embedding is iterated, the communities become so strongly highlighted both in the spatial node arrangement and in the graph structure that it enables HDBSCAN to identify the planted communities at a similar high quality based on both types of inputs.

\bibliography{references}

\begin{thebibliography}{10}
\bibitem{alabprop_code}
\bibinfo{title}{{We used the Python function `asyn\_lpa\_communities', an
  implementation of the asynchronous label propagation algorithm available in
  the `networkx.algorithms.community.label\_propagation' package.}}
\bibitem{ARI_code}
\bibinfo{title}{{We calculated the adjusted Rand index values with the Python
  function `adjusted\_rand\_score' available in the `sklearn.metrics.cluster'
  package.}}
\bibitem{AMI_code}
\bibinfo{title}{{We calculated the adjusted mutual information values with the
  Python function `adjusted\_mutual\_info\_score' available in the
  `sklearn.metrics.cluster' package.}}
\end{thebibliography}

\end{document}